\pdfoutput=1
\documentclass[preprint,11pt]{elsarticle}
\usepackage{geometry}
\usepackage[T1]{fontenc}
\usepackage[utf8]{inputenc}
\usepackage{mathptmx}
\usepackage[scaled=0.90]{helvet}
\usepackage{microtype}
\usepackage{amsmath,amssymb,amsthm,mathtools}
\usepackage{bm}
\usepackage{graphicx,tikz}
\usetikzlibrary{arrows.meta,positioning}
\usepackage{booktabs,array}
\usepackage{xcolor}
\usepackage[font=small,labelfont=bf]{caption}
\usepackage[hidelinks]{hyperref}
\usepackage{orcidlink}
\graphicspath{{./}}

\theoremstyle{plain}
\newtheorem{theorem}{Theorem}
\newtheorem{prop}[theorem]{Proposition}
\newtheorem{corollary}[theorem]{Corollary}
\theoremstyle{remark}
\newtheorem{remark}[theorem]{Remark}

\newcommand{\Cl}{\mathcal{G}}
\newcommand{\phasor}[1]{\vec{#1}}

\journal{Mathematics and Computers in Simulation}
\biboptions{sort&compress}

\begin{document}
\begin{frontmatter}

\title{Symmetrical components in the geometric algebra of \texorpdfstring{$\mathbb{R}^6$}{R6}: modal planes, the discrete Fourier rotor and the conjugate coupling of three-phase operators}
\author[ual]{Francisco G. Montoya\,\orcidlink{0000-0003-4105-565X}\corref{cor1}}
\ead{pagilm@ual.es}
\author[psu]{Ahmad H. Eid}
\cortext[cor1]{Corresponding author.}
\affiliation[ual]{organization={Department of Engineering, University of Almer\'ia},
  city={Almer\'ia},postcode={04120},country={Spain}}
\affiliation[psu]{organization={Department of Electrical Engineering, Faculty of Engineering, Port-Said University},
  city={Port-Fouad},postcode={42526},country={Egypt}}

\begin{abstract}
Modelling and simulation of unbalanced three-phase systems require a distinction between transforming signal coordinates and approximating system operators. We formulate this distinction through Fortescue's symmetrical components in the real geometric algebra $\Cl_{6,0}(\mathbb{R})$. The cyclic phase shift and an orthogonal complex structure jointly determine three modal planes and their sequence projectors. An explicit frame-alignment rotor lifts the realified three-point discrete Fourier transform and exposes its invariant planes and spin periodicity. Every real-linear operator splits orthogonally into complex-linear and conjugate-linear parts. Under cyclic symmetry, the latter exchanges positive and negative sequence while preserving the zero-sequence plane. Its relative Frobenius norm gives the minimum relative error of a complex-linear approximation. Reproducible examples verify equivalent signal representations, harmonic selection and operator diagnostics. A synchronous-frame model derives the mirror-frequency interpretation. A public nonlinear permanent-magnet machine model then tests its practical consequence: under a 2\% negative-sequence voltage disturbance, a compensating voltage calculated with the full local operator reduces total torque-ripple RMS by 96.55\%, compared with 42.39\% using its optimal complex-linear projection. Both voltages are validated in the nonlinear model without retuning. The example connects a specified operator restriction to an observable modelling consequence; a complete classical coupled representation retains the same information.
\end{abstract}

\end{frontmatter}

\section{Introduction}\label{sec:intro}

	The analysis of three-phase electrical systems under unbalanced conditions is a foundational problem in protection, control, harmonic compensation and converter-dominated grids. In the sinusoidal regime considered here, Fortescue's symmetrical components write any phasor triple as the sum of zero-, positive- and negative-sequence parts~\cite{fortescue}. More than a century after its introduction, the same concept remains active in fault analysis, electrical machines, power quality, distributed energy resources and converter control~\cite{chicco2019}.  {Sequence-domain quantities remain the language in which grid requirements are written, while the small-signal behaviour of converter-dominated grids is described by terminal impedance and admittance operators~\cite{harnefors2007complex,rygg2016modified}. This raises a modelling question: which operator couplings are retained, or excluded, when a signal decomposition is used to represent the system dynamics?} Algebraically, Fortescue's decomposition is the spectral decomposition of the cyclic phase-shift operator on $\mathbb{C}^3$. Signal decomposition is therefore universal for phasors; diagonalisation of an impedance or admittance matrix in the same basis is a separate operator-level property that holds only when the system matrix respects the cyclic symmetry.

	An exact change of signal coordinates and a restriction of the operator class have different consequences for a simulation. The former preserves the model, whereas the latter may suppress couplings between its coordinates. A common description of sequence projectors, frame transformations and operator symmetries makes these consequences explicit. In this setting, the error associated with a complex-linear approximation can be separated from the choice of phase or sequence coordinates and then tested in a dynamical example.

	{The classical complex formulation already contains the spectral and metric information. The purpose of a real geometric formulation is to expose how the phase planes, the sequence planes and the invariant planes of the frame transformation fit together, and how operator couplings respect or break the corresponding symmetries. We therefore distinguish equivalences with established linear algebra from the geometric constructions and their consequences.}

	Geometric algebra (GA) provides a real-valued framework that makes the geometry of linear transformations explicit~\cite{hestenes,lasenby2000}. Its electrical-engineering development extends from Bolinder's early treatment~\cite{bolinder1987} to circuit and power-system formulations reviewed in~\cite{chappell2014,montoya2026overview}. Relevant applications include apparent power under non-sinusoidal conditions~\cite{menti2007,castilla2008}, the geometric foundation of the Clarke, Park and FBD transformations~\cite{montoya2022mma}, and the Simple Kirchhoff Rotation framework for multiphase transformations~\cite{eidmontoya2022skr}. Recent work addresses modelling and decoupling control of three-phase systems using $\mathcal{G}_{2,0}$~\cite{velasco2025introducing} and a geometric formulation of Ohm's law for non-symmetric three-phase circuits~\cite{doriacerezo2026ohm}. More broadly, the link between geometric symmetry and executable computations is illustrated by Hrdina \emph{et al.}~\cite{hrdina2021symmetries}, who use Lie-group structure and computer-algebra tools to study a rolling-disc control problem. Here cyclic symmetry and an internal complex structure organise the three-phase operator calculations, while numerical examples verify their consequences for signal decomposition and admittance modelling.

	We develop this connection for Fortescue's symmetrical components and quantify its consequences for admittance modelling. The main contributions are:
		\begin{enumerate}
		
		\item Three-phase phasors are embedded in $\mathbb{R}^6$, where each phase occupies an oriented plane and multiplication by $j$ becomes the simultaneous quarter-turn $\mathcal{J}$ of the three phase planes. The zero-, positive- and negative-sequence components are the orthogonal idempotent projectors $\{\bm{\Pi}_0,\bm{\Pi}_+,\bm{\Pi}_-\}$ onto three further planes, the modal planes, which are jointly determined by the cyclic phase shift and $\mathcal J$; the same space is reached from real measurements by an analytic/Hilbert lift, and the Parseval structure is explicit.
		\item The rotor $\bm{R}\in\mathrm{Spin}(6)$ that carries the phase planes onto the modal planes is obtained in closed form by the characteristic-multivector construction and identified as a spin lift of the realified three-point discrete Fourier transform. Its Cartan angles $0,\pi,\pi/2$ are the Fourier eigenphases, its vector action squared interchanges two phases, the failure of the direct-sum reconstruction formula is traced to the half-turn eigenvalue, and its exponent generates a one-parameter rotor group, the realified fractional Fourier transform, that interpolates between phase and sequence frames, with period eight in the spin group and period four on vectors.
		\item Every real-linear operator on $\mathbb{R}^6$ splits under $\mathcal{J}$ into a complex-linear and a conjugate part. For operators that commute with the cyclic shift the complex-linear part is diagonal in the modal planes and the conjugate part exchanges positive and negative sequence while allowing a zero-sequence diagonal contribution, so the whole inter-sequence coupling of a cyclically symmetric operator is conjugate. The relative weight of the conjugate part, $\rho_{\mathcal J}$, is the minimum relative complex-linear approximation error in the Frobenius norm, which sequence-block norms do not resolve. A synchronous-frame admittance model specifies the frequency channels in which it measures mirror coupling.
		\item Harmonics enter as an orthogonal direct sum $\mathbb{R}^{6h}$ in which any harmonic--sequence component is one idempotent projection, with the classical modulo-three rule as the balanced special case; a sparse $\mathsf{P}/\mathsf{N}$ plane-transfer form of the Clarke-rotated real Fortescue transform is identified; reproducible examples link the signal, harmonic and operator levels.
		
	\end{enumerate}

	For modelling and simulation, the framework connects sequence extraction, harmonic selection and admittance approximation through real linear operators. Five reproducible examples compare classical and geometric signal coordinates, select harmonic--sequence components, distinguish modal and Fourier operator blocks, and evaluate complex/conjugate coupling. A first-order synchronous-frame model connects the algebraic splitting to a physical frequency pair. A nonlinear permanent-magnet synchronous machine (PMSM) then tests how the optimal complex-linear projection changes a torque-ripple compensation calculation. The same identified current--torque map is retained in both designs, and the calculated voltages are applied to the nonlinear model without retuning. These computations distinguish equivalent representations from operator restrictions with measurable consequences. The scripts, numerical records and public-model provenance accompany the manuscript.

		\begin{figure}[htbp]
\centering
\begin{tikzpicture}[
  box/.style={draw=black!55,line width=0.45pt,rounded corners=2pt,
    align=center,text width=4.1cm,minimum height=1.65cm,
    inner xsep=0.2cm,inner ysep=0.12cm,outer sep=0pt,font=\small},
  arrow/.style={-{Stealth[length=1.7mm,width=1.2mm]},
    line width=0.55pt,draw=black!70},
  note/.style={font=\footnotesize,inner sep=2pt,text=black}
]
\node[box,fill=blue!3] (phase) at (2.25,0)
  {\textbf{Phase planes}\\$\bm I_A,\bm I_B,\bm I_C$\\Phasors / Hilbert lift};
\node[box,fill=blue!3] (modal) at (7.75,0)
  {\textbf{Modal planes}\\$\bm B_0,\bm B_+,\bm B_-$\\Joint structure $(\bm S,\mathcal J)$};
\node[box,fill=violet!3] (cartan) at (13.25,0)
  {\textbf{Cartan planes}\\$\bm B_1,\bm B_\pi,\bm B_q$\\Chosen alignment $\bm Q$};
\node[box,fill=blue!3,text width=9.6cm] (conj) at (5,-2.65)
  {\textbf{Complex structure:} $\bm A=\bm A_++\bm A_-$\\
   Sequence blocks $\bm\Pi_k\bm A_\pm\bm\Pi_\ell$\\
   $\rho_{\mathcal J}$: optimal approximation error};
\node[box,fill=violet!3] (sym) at (13.25,-2.65)
  {\textbf{Alignment symmetry}\\
   $\bm Q^2$: phase interchange\\
   $\bm Q$: further frame restriction};

\coordinate (alignment) at (5,0);
\draw[arrow] (phase.east) -- node[note,below=2pt] {$\bm Q$} (modal.west);
\draw[arrow] (alignment) -- (5,1.3)
  -- node[note,above=1pt] {spectrum of $\bm Q$} (13.25,1.3)
  -- (cartan.north);
\fill[black!70] (alignment) circle[radius=0.8pt];

\draw[arrow] (phase.south) -- node[note,right=2pt] {$\mathcal J$}
  (phase.south |- conj.north);
\draw[arrow] (modal.south) -- node[note,right=2pt] {$\bm\Pi_k$}
  (modal.south |- conj.north);
\draw[arrow] (cartan.south) -- node[note,right=2pt] {$\bm\Lambda_\alpha$}
  (sym.north);
\end{tikzpicture}
\caption{{The cyclic shift $\bm S$, complex structure $\mathcal J$ and chosen Fourier alignment $\bm Q$ organise different decompositions. Modal blocks can be combined with the complex/conjugate split; Cartan blocks describe the symmetry of the alignment itself.}}
\label{fig:framework-map}
\end{figure}
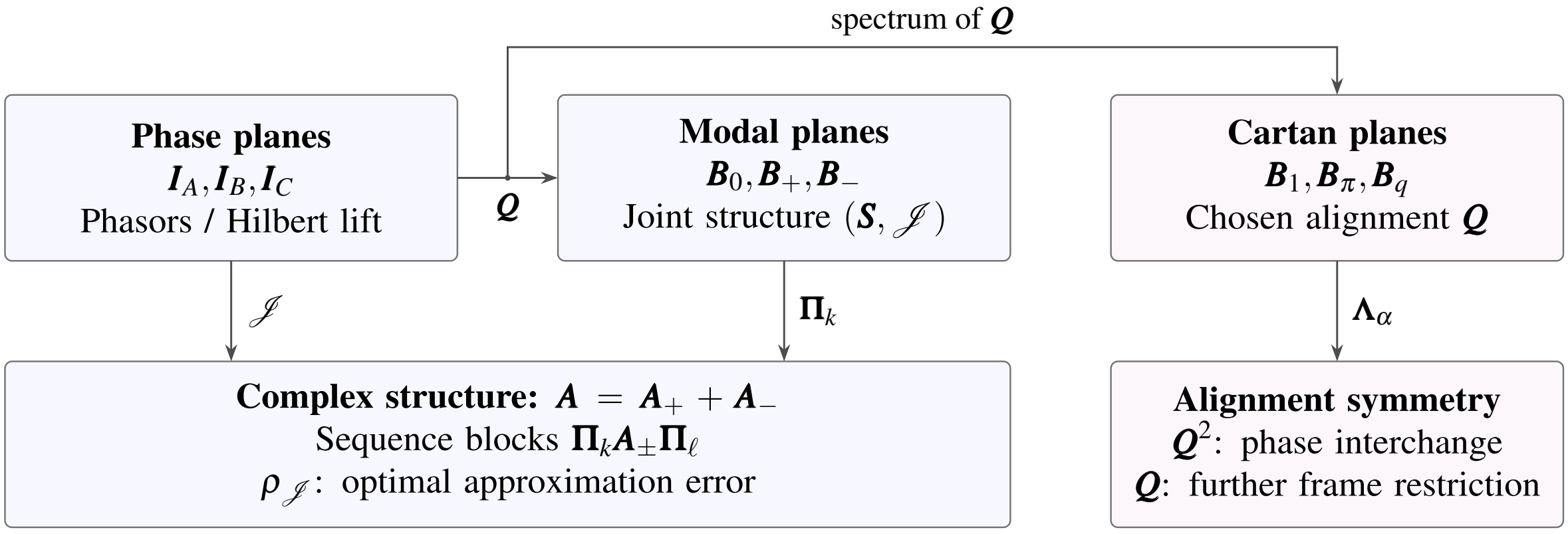

	Figure~\ref{fig:framework-map} connects the two equivalent signal representations to the modal frame and the operator decompositions.
	The remainder of the paper is organised as follows. Section~\ref{sec:framework} introduces the $\mathbb{R}^6$ representation of three-phase phasors and derives the modal basis from the spectral structure of the cyclic shift. Section~\ref{sec:decomposition} develops the projection method, the rotor-based formulation and the equivalence with the classical Fortescue operator. Section~\ref{sec:multiharm} extends the framework to multi-harmonic systems. Sections~\ref{sec:examples}--\ref{sec:conclusion} present worked examples, discussion and conclusions.

	\section{Mathematical Foundation}\label{sec:framework}
	
	\subsection{Six-Dimensional Representation of Three-Phase Systems}
	
	Throughout this paper we work in the sinusoidal regime, in which each phase of a three-phase electrical system at a fixed angular frequency $\omega$ is described by a single complex phasor (or equivalently, by amplitude and phase angle). This is the natural scope of the classical Fortescue construction: each frequency component is represented by its own phasor triple. The extension to multi-harmonic non-sinusoidal signals is treated separately in Section~\ref{sec:multiharm}, where each harmonic order is handled in its own copy of the framework.
	
	In our framework, we represent a three-phase sinusoidal signal in a six-dimensional real space $\mathbb{R}^6$. Each phase (A, B, and C) is associated with two orthogonal basis vectors, one for the cosine component and one for the sine component. Specifically, we define these vectors in terms of the standard Euclidean basis $\{\bm{\sigma}_1, \bm{\sigma}_2, \bm{\sigma}_3, \bm{\sigma}_4, \bm{\sigma}_5, \bm{\sigma}_6\}$ of $\mathbb{R}^6$ as follows:
	
	\begin{equation}
		\begin{aligned}
			\text{Phase A}: &\{\bm{\sigma}_A^c = \bm{\sigma}_1, \bm{\sigma}_A^s = \bm{\sigma}_2\} \\
			\text{Phase B}: &\{\bm{\sigma}_B^c = \bm{\sigma}_3, \bm{\sigma}_B^s = \bm{\sigma}_4\} \\
			\text{Phase C}: &\{\bm{\sigma}_C^c = \bm{\sigma}_5, \bm{\sigma}_C^s = \bm{\sigma}_6\}
		\end{aligned}
	\end{equation}

	These six basis vectors form an orthonormal basis for $\mathbb{R}^6$. That is, they satisfy:
	\begin{align}
		\bm{\sigma}_{i} \cdot \bm{\sigma}_{j} = \delta_{ij}
	\end{align}

	For a three-phase sinusoidal signal with magnitudes $V_a, V_b, V_c$ and phase angles $\theta_A, \theta_B, \theta_C$, we can represent it as a vector $\bm{v} \in \mathbb{R}^6$:
	\begin{equation}
		\begin{aligned}
			\bm{v} &= V_a\cos\theta_A \bm{\sigma}_A^c + V_a\sin\theta_A \bm{\sigma}_A^s + V_b\cos\theta_B \bm{\sigma}_B^c \\
			&+ V_b\sin\theta_B \bm{\sigma}_B^s + V_c\cos\theta_C \bm{\sigma}_C^c + V_c\sin\theta_C \bm{\sigma}_C^s
		\end{aligned}
	\end{equation}

	This representation captures all the information normally contained in complex phasors.
	
	\subsubsection{The embedding \texorpdfstring{$\iota:\mathbb{C}^3\to\mathbb{R}^6$}{iota: C3 to R6}.}\label{sec:emb-injectivity}
	We denote complex phasors with an arrow, writing $\phasor{V}_p=V_p e^{j\theta_p}$; the symbol $V_p$ without arrow denotes the scalar magnitude. The construction above defines a real-linear map
	\begin{equation}\label{eq:iota-def}
		\iota:\mathbb{C}^3 \longrightarrow \mathbb{R}^6,\qquad
		(\phasor{V}_a,\phasor{V}_b,\phasor{V}_c) \longmapsto \bm{v},
	\end{equation}
	with $\bm{v}$ given by the expansion above. The map $\iota$ has three properties that make it the appropriate setting for the present framework. First, $\iota$ is \emph{injective}: any zero output requires $V_p\cos\theta_p = V_p\sin\theta_p = 0$ for each phase $p\in\{a,b,c\}$, which forces $V_p=0$ and hence the input phasor to be zero. Second, $\iota$ is \emph{bijective onto $\mathbb{R}^6$}: any $\bm{v}\in\mathbb{R}^6$ with components $(v_1,\dots,v_6)$ in the canonical basis is the image of the phasor triple $(v_1+j v_2,v_3+j v_4,v_5+j v_6)$. Third, $\iota$ \emph{preserves the real part of the Hermitian inner product}: for any $u,w\in\mathbb{C}^3$,
	\begin{equation}\label{eq:iota-isometry}
		\mathrm{Re}\,\langle u,w\rangle_{\mathbb{C}^3} \;=\; \langle\iota(u),\iota(w)\rangle_{\mathbb{R}^6},
	\end{equation}
	 In particular, $\iota$ is norm-preserving in the sense that $\|\iota(u)\|_{\mathbb{R}^6} = \|u\|_{\mathbb{C}^3}$; the geometric and the complex frameworks carry the same metric information.

	{For a complex-linear matrix $C$, its realification is the operator
$\mathfrak R(C):=\iota\circ C\circ\iota^{-1}$; each entry $a+jb$ becomes the block
$\left[\begin{smallmatrix}a&-b\\b&a\end{smallmatrix}\right]$ in the interleaved coordinate order. A single fixed-frequency copy does not describe transfer between different frequencies.}

	\subsubsection{Analytic lift and internal complex structure.}\label{sec:ahmad-lift}
	The same six-dimensional space can be obtained without starting from complex phasors. Let $\mathcal{H}$ denote the Hilbert transform, with the sign convention $\mathcal{H}\{\cos(\omega t+\theta)\}=\sin(\omega t+\theta)$ for positive-frequency sinusoids. A real phase signal can then be lifted to an oriented Euclidean plane by
	\begin{equation}\label{eq:ahmad-lift}
		u_p(t)
		\;\longmapsto\;
		u_p(t)\,\bm{\sigma}_p^c+\mathcal{H}\{u_p\}(t)\,\bm{\sigma}_p^s,
		\qquad p\in\{A,B,C\}.
	\end{equation}
	For $u_p(t)=V_p\cos(\omega t+\theta_p)$, \eqref{eq:ahmad-lift} is exactly the real vector associated with the analytic signal $V_pe^{j(\omega t+\theta_p)}$. Thus the sine coordinate is not an arbitrary duplication of the measured real signal: it is the quadrature companion that makes phase shifts into orthogonal rotations.

	Throughout the paper, a tilde denotes the Clifford reverse. For a homogeneous product of vectors,
	\[
		\widetilde{\bm{a}_1\bm{a}_2\cdots\bm{a}_r}
		=
		\bm{a}_r\cdots\bm{a}_2\bm{a}_1,
	\]
	extended linearly to arbitrary multivectors. In particular, if $\bm{R}$ is a unit rotor, then $\widetilde{\bm{R}}=\bm{R}^{-1}$ and $\bm{R}\widetilde{\bm{R}}=1$.

	In GA terms, the associated internal complex structure is the simultaneous quarter-turn in the three phase planes. Define the oriented unit phase bivectors
	\[
		\bm{I}_A=\bm{\sigma}_A^c\bm{\sigma}_A^s=\bm{\sigma}_{12},\qquad
		\bm{I}_B=\bm{\sigma}_B^c\bm{\sigma}_B^s=\bm{\sigma}_{34},\qquad
		\bm{I}_C=\bm{\sigma}_C^c\bm{\sigma}_C^s=\bm{\sigma}_{56}.
	\]
	If $\bm{x}=\sum_p\bm{x}_p$ with
	\[
		\bm{x}_p=(\bm{x}\cdot\bm{\sigma}_p^c)\bm{\sigma}_p^c
		+(\bm{x}\cdot\bm{\sigma}_p^s)\bm{\sigma}_p^s,
		\qquad p\in\{A,B,C\},
	\]
	we define the linear map $\mathcal{J}$ by
	\begin{equation}\label{eq:internal-J}
		\mathcal{J}(\bm{x})=\sum_{p\in\{A,B,C\}}\bm{x}_p\bm{I}_p.
	\end{equation}
	Right multiplication by $\bm{I}_p$ is applied only to the component already projected onto the corresponding phase plane, so the result is again a vector. Equivalently,
	\[
		\mathcal{J}(\bm{x})=\bm{R}_{\mathcal{J}}\,\bm{x}\,\widetilde{\bm{R}}_{\mathcal{J}},
		\qquad
		\bm{R}_{\mathcal{J}}=\exp\!\left[-\frac{\pi}{4}(\bm{I}_A+\bm{I}_B+\bm{I}_C)\right].
	\]
	Hence
	\[
		\mathcal{J}(\bm{\sigma}_p^c)=\bm{\sigma}_p^s,\qquad
		\mathcal{J}(\bm{\sigma}_p^s)=-\bm{\sigma}_p^c,\qquad
		\mathcal{J}(\mathcal{J}(\bm{x}))=-\bm{x},
	\]
	and $\mathcal{J}(\bm{u})\cdot\mathcal{J}(\bm{v})=\bm{u}\cdot\bm{v}$. Multiplication by $j$ in the complex phasor picture is precisely the real operation $\mathcal{J}$ after the embedding $\iota$. The phase bivectors $\bm{I}_p$ should not be confused with the modal bivectors $\bm{B}_k$ introduced below; the former generate local Hilbert rotations, while the latter encode Fortescue sequence planes.

	\subsection{Symmetrical Component Basis Vectors}\label{sec:basis-vectors}

	We now construct an orthonormal basis for the symmetrical components (zero, positive, and negative sequence) in this six-dimensional space.

	The classical labels \emph{positive}, \emph{negative} and \emph{zero sequence} refer to the direction in which the magnetic field rotates inside a three-phase machine excited by the corresponding balanced phasor set: a positive-sequence current set produces a rotating field that turns in the direction of normal operation, a negative-sequence set produces a field that rotates in the opposite direction, while a zero-sequence (or homopolar) set consists of three equal, in-phase quantities. Its field depends on the winding arrangement; in an ideal symmetric three-phase winding the fundamental spatial contributions cancel~\cite{fortescue,montoya2026overview}. The terminology dates back to Ferraris's 1895 analysis of single-phase motors as a superposition of two counter-rotating fields, was formalised by Stokvis (1912--1915) and Alexanderson (1913) in the context of phase balancers, and acquired its modern algebraic form with Fortescue's 1918 theorem~\cite{fortescue}. Throughout the paper we use these classical labels; when emphasising the algebraic content rather than the machine-rotational interpretation, we refer to the three sequence subspaces, encoded as $2$-blades, as the \emph{modal bivectors} $\bm{B}_0$, $\bm{B}_+$, $\bm{B}_-$, and to the resulting decomposition as the \emph{geometric modal decomposition} of the three-phase signal.

	\subsubsection{Derivation from the cyclic-shift operator.} Although the explicit formulas below may at first appear ad hoc, they are dictated by a single algebraic principle: the modal basis is the eigenbasis of the cyclic-shift operator on the three-phase signal space. The operator is the linear map that permutes the three phases $A\to B\to C\to A$; in the complex phasor representation, it is the cyclic permutation matrix
	\begin{equation}\label{eq:cyclic-shift-def}
		S \;=\; \begin{pmatrix} 0 & 0 & 1 \\ 1 & 0 & 0 \\ 0 & 1 & 0 \end{pmatrix},
	\end{equation}
	which is unitary and has three distinct eigenvalues $\{1,\phasor{a},\phasor{a}^2\}$ with $\phasor{a}=e^{j2\pi/3}$. The corresponding unit-norm eigenvectors are
	\begin{equation}\label{eq:cyclic-eigenvectors}
		\vec{v}_0 \;=\; \tfrac{1}{\sqrt{3}}\bigl(1,1,1\bigr)^{\!\top},\quad
		\vec{v}_+ \;=\; \tfrac{1}{\sqrt{3}}\bigl(1,\phasor{a}^2,\phasor{a}\bigr)^{\!\top},\quad
		\vec{v}_- \;=\; \tfrac{1}{\sqrt{3}}\bigl(1,\phasor{a},\phasor{a}^2\bigr)^{\!\top},
	\end{equation}
	and are pairwise orthogonal in $\mathbb{C}^3$ because $S$ is unitary with distinct eigenvalues. Under the real embedding $\iota:\mathbb{C}^3\to\mathbb{R}^6$ of Section~\ref{sec:emb-injectivity}, each eigenvector $\vec{v}_k$ produces two real vectors $\iota(\vec{v}_k)$ and $\iota(j\vec{v}_k)$ that, together, span a two-dimensional plane in $\mathbb{R}^6$. In geometric-algebra terms, a plane is naturally encoded as a $2$-blade: we therefore introduce the three \emph{modal bivectors}
	\begin{equation}\label{eq:modal-bivectors}
		\bm{B}_0 \;=\; \iota(\vec{v}_0)\wedge\iota(j\vec{v}_0),\quad
		\bm{B}_+ \;=\; \iota(\vec{v}_+)\wedge\iota(j\vec{v}_+),\quad
		\bm{B}_- \;=\; \iota(\vec{v}_-)\wedge\iota(j\vec{v}_-),
	\end{equation}
	which inherit pairwise orthogonality from the unitarity of $S$ (Section~\ref{sec:ortho-properties}). The analytic-lift viewpoint makes the sine partner explicit:
	\begin{equation}\label{eq:modal-J-invariance}
		\bm{\sigma}_k^s=\mathcal{J}(\bm{\sigma}_k^c),\qquad
		\bm{\Pi}_k\bigl(\mathcal{J}(\bm{x})\bigr)=\mathcal{J}\bigl(\bm{\Pi}_k\bm{x}\bigr),
		\qquad k\in\{0,+,-\},\quad \bm{x}\in\mathbb{R}^6.
	\end{equation}
	Hence every modal plane is invariant under the same internal Hilbert rotation $\mathcal{J}$ that generated the original phase planes. Throughout the paper we use $\bm{B}_k$ both as the bivector and, by the standard GA identification, as the $2$D subspace it represents. The operator that projects onto $\bm{B}_k$ will be denoted $\bm{\Pi}_k$ and is defined in Section~\ref{subsec:spectral}; we reserve different symbols for plane and projector to avoid ambiguity. The explicit formulas in the next three subsubsections are the component expansions of $\iota(\vec{v}_k)$ (cosine vector) and $\iota(j\vec{v}_k)$ (sine vector) in the canonical phase basis of $\mathbb{R}^6$; the factor $1/\sqrt{3}$ enforces unit norm. 

	\subsubsection{Zero Sequence Component}
	
	The cosine and sine basis vectors of the common modal plane $\bm{B}_0$ are the in-phase orthonormal pair:
	
	\begin{equation}
		\begin{aligned}
			\bm{\sigma}_0^c &= \frac{1}{\sqrt{3}}(\bm{\sigma}_A^c + \bm{\sigma}_B^c + \bm{\sigma}_C^c) \\
			\bm{\sigma}_0^s &= \frac{1}{\sqrt{3}}(\bm{\sigma}_A^s + \bm{\sigma}_B^s + \bm{\sigma}_C^s)
		\end{aligned}
	\end{equation}

	Together $\bm{\sigma}_0^c$ and $\bm{\sigma}_0^s$ span the common modal plane $\bm{B}_0$, which is the image under the embedding $\iota$ of the one-dimensional complex eigenspace of the cyclic shift associated with the eigenvalue $1$.

	\subsubsection{Positive Sequence Component}
	
	The cosine and sine basis vectors of the forward modal plane $\bm{B}_+$ are constructed as:
	\begin{equation}
		\begin{aligned}
			\bm{\sigma}_+^c &= \frac{1}{\sqrt{3}}\left(\bm{\sigma}_A^c - \frac{1}{2}\bm{\sigma}_B^c - \frac{\sqrt{3}}{2}\bm{\sigma}_B^s - \frac{1}{2}\bm{\sigma}_C^c + \frac{\sqrt{3}}{2}\bm{\sigma}_C^s\right) \\
			\bm{\sigma}_+^s &= \frac{1}{\sqrt{3}}\left(\bm{\sigma}_A^s + \frac{\sqrt{3}}{2}\bm{\sigma}_B^c - \frac{1}{2}\bm{\sigma}_B^s - \frac{\sqrt{3}}{2}\bm{\sigma}_C^c - \frac{1}{2}\bm{\sigma}_C^s\right)
		\end{aligned}
	\end{equation}
	
	\subsubsection{Negative Sequence Component}
	
	The cosine and sine basis vectors of the backward modal plane $\bm{B}_-$ are constructed as:
	\begin{equation}
		\begin{aligned}
			\bm{\sigma}_-^c &= \frac{1}{\sqrt{3}}\left(\bm{\sigma}_A^c - \frac{1}{2}\bm{\sigma}_B^c + \frac{\sqrt{3}}{2}\bm{\sigma}_B^s - \frac{1}{2}\bm{\sigma}_C^c - \frac{\sqrt{3}}{2}\bm{\sigma}_C^s\right) \\
			\bm{\sigma}_-^s &= \frac{1}{\sqrt{3}}\left(\bm{\sigma}_A^s - \frac{\sqrt{3}}{2}\bm{\sigma}_B^c - \frac{1}{2}\bm{\sigma}_B^s + \frac{\sqrt{3}}{2}\bm{\sigma}_C^c - \frac{1}{2}\bm{\sigma}_C^s\right)
		\end{aligned}
	\end{equation}
	
	\subsection{Orthogonality Properties}\label{sec:ortho-properties}

	The modal basis is orthonormal because it is obtained from the unitary eigenbasis of the cyclic-shift matrix through the isometric real embedding $\iota$. Equivalently, for $k,\ell\in\{0,+,-\}$,
	\begin{equation}
		\bm{\sigma}_k^c\cdot\bm{\sigma}_\ell^c = \delta_{k\ell},\quad
		\bm{\sigma}_k^s\cdot\bm{\sigma}_\ell^s = \delta_{k\ell},\quad
		\bm{\sigma}_k^c\cdot\bm{\sigma}_\ell^s = 0,
	\end{equation}
	and the three sequence planes form an orthogonal direct sum of $\mathbb{R}^6$. Consequently the modal bivectors are mutually orthogonal:
	\begin{equation}\label{eq:bivector-inner-product}
		\bm{B}_k\cdot\bm{B}_\ell=0,\qquad k\neq \ell.
	\end{equation}
	This is the geometric reason that sequence energies add independently in the Parseval identities below. Component-wise verification of the displayed identities is included in the accompanying scripts.

	\subsection{Spectral resolution via orthogonal idempotent projectors}
	\label{subsec:spectral}

	The decomposition of $\mathbb{R}^6$ into the three modal planes admits a clean algebraic reformulation in terms of \emph{orthogonal idempotent projectors} acting on $\mathbb{R}^6$. For each modal label $k\in\{0,+,-\}$, define the linear operator $\bm{\Pi}_k$ on $\mathbb{R}^6$ by
	\begin{equation}\label{eq:projector-def}
		\bm{\Pi}_k\,\bm{v}\;=\;(\bm{v}\cdot\bm{\sigma}_k^c)\,\bm{\sigma}_k^c \;+\;(\bm{v}\cdot\bm{\sigma}_k^s)\,\bm{\sigma}_k^s.
	\end{equation}
	This is the orthogonal projector onto the modal plane spanned by $\{\bm{\sigma}_k^c,\bm{\sigma}_k^s\}$ and admits the equivalent geometric-algebra expression
	\begin{equation}\label{eq:projector-bivector}
		\bm{\Pi}_k\,\bm{v} \;=\; (\bm{v}\,\rfloor\,\bm{B}_k)\,\bm{B}_k^{-1},
	\end{equation}
	where $\bm{B}_k = \bm{\sigma}_k^c\wedge\bm{\sigma}_k^s$ is the modal bivector itself, equivalently expressed in the modal basis, and $\bm{B}_k^{-1} = -\bm{B}_k$ since $\bm{B}_k^{\,2} = -1$. Equation~\eqref{eq:projector-def} uses modal coordinates, while \eqref{eq:projector-bivector} expresses the same projection intrinsically through the bivector.

	\subsubsection{Complete system of orthogonal idempotents.} The family $\{\bm{\Pi}_0,\bm{\Pi}_+,\bm{\Pi}_-\}$ satisfies
	\begin{equation}\label{eq:idemp-system}
	\bm{\Pi}_k^2 = \bm{\Pi}_k,\qquad
	\bm{\Pi}_k\bm{\Pi}_\ell = 0 \ \text{for } k\neq \ell,\qquad
	\bm{\Pi}_0 + \bm{\Pi}_+ + \bm{\Pi}_- = \mathsf{1}_{\mathrm{op}},
	\end{equation}
	where $\mathsf{1}_{\mathrm{op}}$ denotes the identity operator on vectors in $\mathbb{R}^6$. It is an operator symbol, not a Clifford multivector; the scalar unit of $\Cl_{6,0}(\mathbb{R})$ remains $1$, and the pseudoscalar is $\bm{\sigma}_1\bm{\sigma}_2\bm{\sigma}_3\bm{\sigma}_4\bm{\sigma}_5\bm{\sigma}_6$. The idempotence $\bm{\Pi}_k^2=\bm{\Pi}_k$ and the mutual orthogonality $\bm{\Pi}_k\bm{\Pi}_\ell=0$ follow directly from the orthonormality of the modal basis (Section~\ref{sec:ortho-properties}); the completeness $\sum_k\bm{\Pi}_k=\mathsf{1}_{\mathrm{op}}$ expresses that the three modal planes $\bm{B}_0,\bm{B}_+,\bm{B}_-$ form an orthogonal direct-sum decomposition of $\mathbb{R}^6$.

\subsubsection{Joint structure of the cyclic shift and complex structure}
Writing $\bm S=\mathfrak R(S)$ and $I=\mathsf1_{\mathrm{op}}$, the same projectors have the real polynomial form
\begin{equation}\label{eq:joint-projectors}
\bm\Pi_0=\frac{I+\bm S+\bm S^2}{3},\qquad
\bm\Pi_\pm=\frac{I-\bm\Pi_0}{2}
\mp\frac{\mathcal J(\bm S-\bm S^2)}{2\sqrt3}.
\end{equation}
Indeed, $\bm S-\bm S^2$ acts as $\pm\sqrt3\,\mathcal J$ on $\bm B_\pm$ and vanishes on $\bm B_0$. Over $\mathbb R$, the two nonzero-sequence planes are equivalent cyclic representations, so $\bm S$ alone does not distinguish them canonically. The specified $\mathcal J$ resolves that ambiguity. Equation~\eqref{eq:joint-projectors} connects the modal projectors to the two commuting structures without selecting local coordinates.

\subsubsection{Reformulation of the decomposition.} The Fortescue-style decomposition $\bm{v}=\bm{v}_0+\bm{v}_++\bm{v}_-$ now reads as the spectral identity
	\begin{equation}\label{eq:spectral-id}
		\bm{v} \;=\; \bm{\Pi}_0\bm{v} + \bm{\Pi}_+\bm{v} + \bm{\Pi}_-\bm{v},\qquad
		\bm{v}_k\;:=\;\bm{\Pi}_k\bm{v}.
	\end{equation}
	Section~\ref{sec:projection-method} gives the corresponding local coordinates for interfacing with classical phasors. %

	\section{Decomposition into Symmetrical Components}\label{sec:decomposition}
	
	\subsection{Projection method}\label{sec:projection-method}

	The decomposition of a three-phase signal $\bm{v}\in\mathbb{R}^6$ into its symmetrical components is the spectral identity \eqref{eq:spectral-id} with the modal projectors $\bm{\Pi}_k$ of \eqref{eq:projector-def}--\eqref{eq:projector-bivector}: the $k$-th sequence component is the bivector-contraction
	\begin{equation}\label{eq:projection-summary}
		\bm{v}_k \;=\; \bm{\Pi}_k\bm{v} \;=\; (\bm{v}\,\rfloor\,\bm{B}_k)\,\bm{B}_k^{-1},\qquad k\in\{0,+,-\},
	\end{equation}
	which computes the orthogonal projection of $\bm{v}$ onto the modal bivector $\bm{B}_k$ intrinsically, without recourse to scalar dot products. When the two scalar coordinates of $\bm{v}_k$ in the local basis $\{\bm{\sigma}_k^c,\bm{\sigma}_k^s\}$ are needed explicitly --- for instance, to interface with a classical phasor representation --- they are recovered by inner products,
	\begin{equation}\label{eq:coord-extraction}
		V_k^c \;=\; \bm{v}\cdot\bm{\sigma}_k^c,\qquad V_k^s \;=\; \bm{v}\cdot\bm{\sigma}_k^s,
	\end{equation}

	\subsection{Rotor-based formulation}\label{sec:rotor-method}

	An alternative to projection-based decomposition is to apply a single rotor $\bm{R}\in\mathrm{Spin}(6)$ that aligns the canonical basis $\{\bm{\sigma}_1,\dots,\bm{\sigma}_6\}$ of $\mathbb{R}^6$ with the modal basis $\{\bm{\sigma}_0^c,\bm{\sigma}_0^s,\bm{\sigma}_+^c,\bm{\sigma}_+^s,\bm{\sigma}_-^c,\bm{\sigma}_-^s\}$. Under this rotation, the modal coordinates of any signal $\bm{v}$ are read directly as the components of the rotated vector in the canonical basis, and the modal projectors transform into the canonical pair-of-axes projectors of $\mathbb{R}^6$. The construction is fully reversible and isometric.

	For this fixed ordered modal frame, the rotor $\bm{R}$ exists and is unique up to the global $\pm$ sign of the spin double cover $\mathrm{Spin}(6)\to\mathrm{SO}(6)$, because the $6\times 6$ matrix whose columns are the modal basis vectors lies in $\mathrm{SO}(6)$. Its explicit construction in dimension six does not reduce to the elementary single-plane rotor formula $(1+\bm{B}\bm{A}^{\dagger})/\|1+\bm{B}\bm{A}^{\dagger}\|$, because the modal alignment carries more than one independent invariant plane; the construction in Section~\ref{subsubsec:rotor-charmv} handles these planes jointly. We first specify the diagonalisation target and frame convention.

Write $\bm Q=\mathfrak R(\bm F_3)$ for this ordered frame matrix, with $\bm F_3$ defined in Section~\ref{subsubsec:dft}. Its action is $\bm Q\bm x=\bm R\bm x\widetilde{\bm R}$, and modal coordinates are obtained by $\bm Q^\top\bm v$.

\subsubsection{Simultaneous diagonalisation and frame convention}
\label{subsubsec:rotor-diag}
Let $\{\bm{E}_0,\bm{E}_+,\bm{E}_-\}$ denote the canonical \emph{pair-of-axes} projectors of $\mathbb{R}^6$ associated with three disjoint $2$-blocks of coordinates,
	\begin{equation}
		\begin{aligned}
		\bm{E}_0\bm{v}
		&=(\bm{v}\cdot\bm{\sigma}_1)\bm{\sigma}_1
		+(\bm{v}\cdot\bm{\sigma}_2)\bm{\sigma}_2,\\
		\bm{E}_+\bm{v}
		&=(\bm{v}\cdot\bm{\sigma}_3)\bm{\sigma}_3
		+(\bm{v}\cdot\bm{\sigma}_4)\bm{\sigma}_4,\\
		\bm{E}_-\bm{v}
		&=(\bm{v}\cdot\bm{\sigma}_5)\bm{\sigma}_5
		+(\bm{v}\cdot\bm{\sigma}_6)\bm{\sigma}_6.
		\end{aligned}
	\end{equation}
	Each $\bm{E}_k$ is the orthogonal projector onto the corresponding canonical $2$-plane of $\mathbb{R}^6$, and the family $\{\bm{E}_0,\bm{E}_+,\bm{E}_-\}$ satisfies exactly the same algebraic identities \eqref{eq:idemp-system} as the modal projectors $\{\bm{\Pi}_0,\bm{\Pi}_+,\bm{\Pi}_-\}$.

	Consequently, the inverse frame change simultaneously diagonalises the three modal idempotents:
\begin{equation}\label{eq:rotor-diag}
\bm Q^\top\bm\Pi_k\bm Q=\bm E_k,\qquad k\in\{0,+,-\}.
\end{equation}
This is an identity between linear operators, distinct from the Clifford sandwich defining $\bm Q$. Simultaneous diagonalisation determines the modal planes but allows rotations of each local frame; it does not by itself determine a unique rotor. The ordered eigenvectors in \eqref{eq:cyclic-eigenvectors} fix that freedom and hence $\bm R$ up to sign. Its Cartan data refer to this Fourier frame convention.

	\subsubsection{Closed form via characteristic multivectors}
	\label{subsubsec:rotor-charmv}

	Concretely, $\bm{R}$ is obtained from the characteristic-multivector theorem of Lasenby, Lasenby and Matsantonis~\cite{lasenby2024rotor}, in the seeded special-case form of \cite[\S3.1]{lasenby2024rotor} --- the rotor-reconstruction version of the averaging method introduced by Shirokov for the spin groups~\cite{shirokov2019averaging} --- which is required here because the direct sum of the canonical and modal frames vanishes for this alignment. The complete derivation is carried out in the Supplementary Material (Section~S1); this subsection states the result. Label the target modal basis as
	\[
	\{\bm{u}_1,\dots,\bm{u}_6\} = \{\bm{\sigma}_0^c,\bm{\sigma}_0^s,\bm{\sigma}_+^c,\bm{\sigma}_+^s,\bm{\sigma}_-^c,\bm{\sigma}_-^s\},
	\]
	and introduce the conjugate constants
	\[
		\rho_-:=\sqrt{3}-1,\qquad \rho_+:=\sqrt{3}+1,
	\]
	so that $\rho_+\rho_-=2$, $1/\rho_-=\rho_+/2$, and $3\pm\sqrt{3}=\sqrt{3}\rho_\pm$. The rotor is written most compactly through two oriented planes:
	\begin{align}\label{eq:LS-reduced-planes}
		\bm{x}_\pi &= \tfrac{1}{\sqrt{\sqrt{3}\rho_+}}\bigl(\bm{\sigma}_1-\tfrac{\rho_+}{2}(\bm{\sigma}_3+\bm{\sigma}_5)\bigr),
		& \bm{y}_\pi &= \tfrac{1}{\sqrt{\sqrt{3}\rho_+}}\bigl(\bm{\sigma}_2-\tfrac{\rho_+}{2}(\bm{\sigma}_4+\bm{\sigma}_6)\bigr),\notag\\
		\bm{x}_q &= \tfrac{1}{\sqrt{2}}\bigl(\bm{\sigma}_3-\bm{\sigma}_5\bigr),
		& \bm{y}_q &= \tfrac{1}{\sqrt{2}}\bigl(\bm{\sigma}_4-\bm{\sigma}_6\bigr).
	\end{align}
	Define the corresponding unit bivectors
	\begin{equation}\label{eq:LS-reduced-bivectors}
		\bm{B}_\pi:=\bm{x}_\pi\bm{y}_\pi,\qquad
		\bm{B}_q:=\bm{x}_q\bm{y}_q,
		\qquad
		\bm{B}_\pi^2=\bm{B}_q^2=-1.
	\end{equation}
	The characteristic-multivector construction then returns the unit rotor in the closed form
	\begin{equation}\label{eq:R-closed-form-T0}
		{\;
		\bm{R}
		\;=\;
		\frac{1+\bm{B}_q}{\sqrt{2}}\,\bm{B}_\pi,\;}
	\end{equation}
	with the frame convention above. It contains only grades $\{2,4\}$ of $\Cl_{6,0}^+(\mathbb{R})$; in particular it has zero scalar part, which is the algebraic reason the general-position direct-sum formula of~\cite{lasenby2024rotor} degenerates for this frame pair and the seeded form of the construction is needed (Supplementary Material, Proposition~S1). The geometric meaning of $\bm{B}_\pi$ and $\bm{B}_q$ is not assumed at this stage; it is identified in the following subsection, where they are shown to be two Cartan eigenblades of the Fortescue alignment.

	\subsubsection{Cartan interpretation of the characteristic rotor}
	\label{subsubsec:rotor-cartan}

	We now identify the invariant planes of the rotor in \eqref{eq:R-closed-form-T0}. Spectral diagonalisation of the associated real orthogonal map $\bm{Q}\bm{x}:=\bm{R}\bm{x}\widetilde{\bm{R}}$ shows that its eigenvalues are $\{1,-1,-j\}$, each paired with its complex conjugate; correspondingly $\bm{R}$ acts on three mutually orthogonal $2$-planes with rotation angles $0$, $\pi$, and $\pi/2$. The half-turn and quadrature planes are precisely the two planes already introduced in \eqref{eq:LS-reduced-planes}. The remaining fixed plane is
	\begin{align}\label{eq:cartan-eigenframes}
		\bm{x}_1 &= \tfrac{1}{\sqrt{\sqrt{3}\rho_-}}\bigl(\bm{\sigma}_1+\tfrac{\rho_-}{2}(\bm{\sigma}_3+\bm{\sigma}_5)\bigr),
		& \bm{y}_1 &= \tfrac{1}{\sqrt{\sqrt{3}\rho_-}}\bigl(\bm{\sigma}_2+\tfrac{\rho_-}{2}(\bm{\sigma}_4+\bm{\sigma}_6)\bigr).
	\end{align}
	Thus the Cartan frame is the ordered triple of planes
	\[
		(\bm{x}_1,\bm{y}_1),\qquad
		(\bm{x}_\pi,\bm{y}_\pi),\qquad
		(\bm{x}_q,\bm{y}_q),
	\]
	where the last two pairs are those of \eqref{eq:LS-reduced-planes}. The three pairs are individually orthonormal, and the three planes are mutually orthogonal in $\mathbb{R}^6$. They are also compatible with the analytic lift:
	\begin{equation}\label{eq:cartan-J-invariance}
		\bm{y}_1=\mathcal{J}(\bm{x}_1),\qquad
		\bm{y}_\pi=\mathcal{J}(\bm{x}_\pi),\qquad
		\bm{y}_q=\mathcal{J}(\bm{x}_q).
	\end{equation}
	Thus the Cartan eigenplanes are not detached from the Hilbert/phasor structure: they are $\mathcal{J}$-invariant real planes, i.e. real representatives of complex one-dimensional eigenspaces of the alignment operator. Introducing the fixed-sector blade
	\begin{equation}\label{eq:Bpi-Bq}
		\bm{B}_1:=\bm{x}_1\bm{y}_1,
		\qquad
		\bm{B}_1^2=\bm{B}_\pi^2=\bm{B}_q^2=-1,
	\end{equation}
	with $\bm{B}_\pi$ and $\bm{B}_q$ as in \eqref{eq:LS-reduced-bivectors},
	the induced exterior action satisfies
	\begin{equation}\label{eq:eigenblade-action}
		(\bigwedge\nolimits^2 \bm{Q})\,\bm{B}_\alpha
		\;=\;
		\bm{B}_\alpha,
		\qquad \alpha\in\{1,\pi,q\}.
	\end{equation}
	In this precise sense $\bm{B}_1$, $\bm{B}_\pi$ and $\bm{B}_q$ are the real \emph{eigenblades} of the Fortescue alignment. This follows the eigenblade interpretation used in the Simple Kirchhoff Rotation framework~\cite{eidmontoya2022skr}: a conjugate pair of complex eigenvectors of a real orthogonal map is replaced by a single real oriented plane. The blade records the invariant plane, while the associated planar angle records the non-trivial spectral phase. The rotation data are shown in Table~\ref{tab:eigenblade}, up to the orientation convention chosen for each blade.

\begin{table}[htbp]
	\centering
	\caption{Eigenblades and rotation angles by spectral sector.}
	\label{tab:eigenblade}
	\begin{tabular}{lccc}
		\toprule
		Sector & Eigenblade & Vector eigenvalues of $\bm{Q}$ & Rotation angle \\
		\midrule
		Fixed      & $\bm{B}_1$   & $1,\ 1$                & $0$     \\
		Half-turn  & $\bm{B}_\pi$ & $-1,\ -1$              & $\pi$   \\
		Quadrature & $\bm{B}_q$   & $e^{\pm j\pi/2}=\pm j$ & $\pi/2$ \\
		\bottomrule
	\end{tabular}
\end{table}

	Mutual orthogonality of the supporting planes implies the commutator identity
	\begin{equation}\label{eq:commuting-bivectors}
		[\bm{B}_\pi,\bm{B}_q] \;=\; 0.
	\end{equation}
	The factorisation obtained from the characteristic multivector is therefore precisely the Cartan form
	\begin{equation}\label{eq:R-cartan-product}
		{\;\bm{R} \;=\; \exp(0\,\bm{B}_1)\,\bm{R}_q\,\bm{R}_\pi \;=\; \frac{1+\bm{B}_q}{\sqrt{2}}\;\bm{B}_\pi,\;}
	\end{equation}
	where $\exp(0\,\bm{B}_1)$ is the trivial factor associated with the fixed eigenblade, $\bm{R}_q=(1+\bm{B}_q)/\sqrt{2}=\exp(\tfrac{\pi}{4}\bm{B}_q)$ is the rotor producing a vector rotation of angle $\pi/2$ in the plane $\bm{B}_q$, and $\bm{R}_\pi=\bm{B}_\pi=\exp(\tfrac{\pi}{2}\bm{B}_\pi)$ is the rotor producing a vector rotation of angle $\pi$ in the plane $\bm{B}_\pi$. Equivalently, by \eqref{eq:commuting-bivectors}, the rotor is a single exponential of a (non-simple) bivector,
	\begin{equation}\label{eq:R-single-exp}
		\bm{R} \;=\; \exp\!\left(\tfrac{\pi}{4}\,\bm{B}_q \;+\; \tfrac{\pi}{2}\,\bm{B}_\pi\right).
	\end{equation}
	This is the Cartan canonical interpretation of the characteristic rotor: the (essentially unique) decomposition of an element of $\mathrm{SO}(6)$ as a product of at most three commuting plane rotations, here reduced to two non-trivial factors since the first eigenplane is fixed.

	\subsubsection{Identification with the discrete Fourier transform}
	\label{subsubsec:dft}

	The Cartan data of Table~\ref{tab:eigenblade} are not specific to the present construction: they are the spectral data of the discrete Fourier transform. Write $\omega=e^{-j2\pi/3}=\phasor{a}^2$ and let
	\begin{equation}\label{eq:dft3}
		\bm{F}_3=\frac{1}{\sqrt3}\bigl[\omega^{mn}\bigr]_{m,n=0}^{2}
		=\frac{1}{\sqrt3}\begin{pmatrix}1&1&1\\ 1&\phasor{a}^2&\phasor{a}\\ 1&\phasor{a}&\phasor{a}^2\end{pmatrix}
	\end{equation}
	be the unitary three-point discrete Fourier transform. Its columns are the eigenvectors \eqref{eq:cyclic-eigenvectors} of the cyclic shift, so the modal-frame matrix is $\bm U=\bm Q=\mathfrak R(\bm F_3)$, with paired columns $\iota(v_k),\iota(jv_k)$, and the rotor is a spin lift of this real orthogonal map. Everything stated about $\bm{R}$ in the two previous subsections can therefore be read against the classical eigenstructure of the Fourier matrix.

	\begin{prop}\label{prop:dft}
	Let $\bm{Q}\bm{x}=\bm{R}\bm{x}\widetilde{\bm{R}}$ be the frame-alignment map of Section~\ref{subsubsec:rotor-diag}. Then:
	\begin{enumerate}
		\item $\bm{Q}$ is the realification of $\bm{F}_3$. The eigenvalues of $\bm{F}_3$ are $1$, $-1$ and $-j$, each simple, in agreement with the multiplicities $\lfloor N/4\rfloor+1$, $\lfloor (N+2)/4\rfloor$, $\lfloor (N+1)/4\rfloor$ and $\lfloor (N-1)/4\rfloor$ of the eigenvalues $1,-1,-j,j$ of the $N$-point transform~\cite{mcclellan1972}, evaluated at $N=3$. The Cartan angles $0,\pi,\pi/2$ are the Fourier eigenphases, and the eigenblades $\bm{B}_1,\bm{B}_\pi,\bm{B}_q$ are the realified Fourier eigenvectors $(1,\rho_-/2,\rho_-/2)$, $(1,-\rho_+/2,-\rho_+/2)$ and $(0,1,-1)$.
		\item $\bm{F}_3^{\,2}$ is the permutation that exchanges the second and third coordinates, so $\bm{Q}^2$ is the realified interchange of phases $b$ and $c$; it fixes $\bm{B}_0$ and maps $\bm{B}_+$ onto $\bm{B}_-$. Consequently $\bm{\Lambda}_1+\bm{\Lambda}_\pi=\tfrac12(\mathsf{1}_{\mathrm{op}}+\bm{Q}^2)$ is the projector onto the subspace symmetric under that interchange and $\bm{\Lambda}_q=\tfrac12(\mathsf{1}_{\mathrm{op}}-\bm{Q}^2)$ the projector onto the antisymmetric one, and $\bm{Q}^4=\mathsf{1}_{\mathrm{op}}$ is the classical $\bm{F}_3^{\,4}=\bm{I}_3$.
		\item The scalar part of a product of commuting plane rotors with angles $\theta_k$ is $\prod_k\cos(\theta_k/2)$. For the Fourier frame it vanishes because of the half-turn, which is the algebraic origin of the degenerate direct sum $\bm{M}=0$ in the characteristic-multivector construction. Since $-1$ is an eigenvalue of the $N$-point Fourier matrix for every $N\geq 2$, the direct-sum formula degenerates for Fourier frames in every dimension and the seeded construction is always required.
	\end{enumerate}
	\end{prop}

	\begin{proof}
	The first statement is the observation preceding the proposition together with the eigenvector computation $\bm{F}_3(1,c,c)^{\!\top}=\tfrac{1}{\sqrt3}(1+2c,\,1-c,\,1-c)^{\!\top}$, which is an eigenvector with eigenvalue $\lambda\in\{1,-1\}$ exactly when $1+2c=\sqrt3\lambda$ and $1-c=\sqrt3\lambda c$, i.e.\ $c=\rho_-/2$ for $\lambda=1$ and $c=-\rho_+/2$ for $\lambda=-1$, while $\bm{F}_3(0,1,-1)^{\!\top}=\tfrac{1}{\sqrt3}(0,\omega-\omega^2,\omega^2-\omega)^{\!\top}=-j\,(0,1,-1)^{\!\top}$. For the second, $(\bm{F}_3^{\,2})_{mn}=\tfrac13\sum_k\omega^{k(m+n)}=\delta_{m+n\equiv 0\ (\mathrm{mod}\ 3)}$, the reversal permutation, which fixes the first coordinate and exchanges the other two; the images of the modal planes follow from $\bm{v}_+=\tfrac{1}{\sqrt3}(1,\phasor{a}^2,\phasor{a})^{\!\top}\mapsto\bm{v}_-$. Expanding $\prod_k(\cos\tfrac{\theta_k}{2}+\sin\tfrac{\theta_k}{2}\bm{B}_k)$ for mutually orthogonal commuting blades gives the scalar part in the third statement; the multiplicity of the eigenvalue $-1$ is $\lfloor (N+2)/4\rfloor\geq1$ for $N\geq2$~\cite{mcclellan1972}.
	\end{proof}

	For complex-linear operators, commuting with $\bm Q$ is equivalent to diagonality in the complex Fourier eigenbasis~\cite{dickinson1982}; the intersection with the complex circulants consists only of scalar matrices. For arbitrary real-linear operators the commutant is larger: its restrictions to the fixed and half-turn planes are arbitrary real $2\times2$ blocks, while the quadrature block has the form $aI+b\mathcal J$. Its real dimension is therefore $4+4+2=10$, rather than the six dimensions of the complex-linear subspace.

The trace of $\bm F_3$ is the quadratic Gauss sum
$\tfrac{1}{\sqrt3}\sum_{k=0}^{2}e^{-j2\pi k^2/3}=-j$.
Its vanishing real part gives $\operatorname{tr}\bm Q=2\sum_k\cos\theta_k=0$, consistently with the Cartan angles.

\begin{prop}[Fractional Fourier path]\label{prop:fractional}
Set $\bm B_R=\tfrac\pi4\bm B_q+\tfrac\pi2\bm B_\pi$,
$\bm R(t)=\exp(t\bm B_R)$ and $\bm Q(t)\bm x=\bm R(t)\bm x\widetilde{\bm R(t)}$.
Then $\bm R(t)$ is a one-parameter subgroup of $\mathrm{Spin}(6)$, with
\begin{equation}\label{eq:spin-period}
\bm R(1)=\bm R,\quad \bm R(2)=-\bm B_q,\quad
\bm R(4)=-1,\quad \bm R(8)=1.
\end{equation}
On vectors, $\bm Q(2)$ interchanges phases $b,c$ and $\bm Q(4)=I$.
Every $\bm Q(t)$ commutes with $\mathcal J$ and equals $\mathfrak R(\bm F_3^{\,t})$, where the eigenphase branch $(0,-\pi/2,-\pi)$ gives
\begin{equation}\label{eq:fractional-dft}
\bm F_3^{\,t}=\bm P_1+e^{-j\pi t/2}\bm P_{-j}+e^{-j\pi t}\bm P_{-1}.
\end{equation}
Here $\bm P_\lambda$ are the complex spectral projectors of $\bm F_3$. For $N=3$ this is the discrete fractional Fourier transform of~\cite{candan2000}.
\end{prop}
\begin{proof}
The two bivectors commute and square to $-1$, so
$\bm R(2)=\exp(\tfrac\pi2\bm B_q)\exp(\pi\bm B_\pi)=-\bm B_q$;
its square is $-1$. Thus the spin lift has period eight while its vector action has period four. Each Cartan plane is $\mathcal J$-invariant, and $\exp(t\theta\bm B/2)$ induces complex multiplication by $e^{-jt\theta}$ on its oriented complex line. This gives \eqref{eq:fractional-dft}; the simple eigenvalues at $N=3$ have indices $0,1,2$ in~\cite{candan2000}.
\end{proof}
The frames $\{\bm Q(t)\bm\sigma_i\}$ interpolate between phase and modal frames, and the projectors $\bm Q(t)\bm E_k\bm Q(t)^\top$ equal $\bm\Pi_k$ at $t=1$. The chosen logarithm fixes this path; other eigenphase branches can give different interpolations with the same endpoint.

	\begin{remark}\label{rem:eigen-vs-modal}
	The three eigenplanes assembled from \eqref{eq:LS-reduced-planes} and \eqref{eq:cartan-eigenframes} are \emph{not} the modal planes $\bm{B}_0,\bm{B}_+,\bm{B}_-$ of Section~\ref{subsec:spectral}. The fixed eigenplane is spanned by $(\bm{x}_1,\bm{y}_1)$ with canonical coefficients in ratio $1:\rho_-/2:\rho_-/2$, whereas the modal plane $\bm{B}_0$ is spanned by $(\bm{\sigma}_0^c,\bm{\sigma}_0^s)$ with coefficients in ratio $1:1:1$: the two pairs of planes generate distinct two-dimensional subspaces of $\mathbb{R}^6$. The modal planes are the \emph{image} planes that carry the symmetrical components after the rotation, whereas the eigenplanes are the \emph{invariant} planes of the rotation itself. Concretely, $\bm{R}$ rotates each canonical pair-of-axes plane $\bm{E}_k$ into the modal plane $\bm{B}_k$, while leaving its own eigenplanes setwise invariant. The two viewpoints are connected by $\bm{R}$ but cannot be identified at the level of subspaces.
	\end{remark}

	\subsubsection{A second spectral resolution and operator blocks}
	\label{subsubsec:fortescue-operator-blocks}

	Recall that $\bm{Q}\in\mathrm{SO}(6)$ is the real orthogonal operator implemented by the rotor,
	\begin{equation}
		\bm{Q}\bm{x} \;:=\; \bm{R}\bm{x}\widetilde{\bm{R}},
		\qquad \bm{Q}\bm{\sigma}_i=\bm{u}_i.
	\end{equation}
	The Cartan decomposition above shows that $\bm{Q}$ has order four: $\bm{Q}^4=\mathsf{1}_{\mathrm{op}}$. Equivalently, its minimal polynomial splits into the real factors associated with eigenvalues $1$, $-1$ and the quadrature pair $\pm j$. Consequently its three invariant plane projectors can be written directly as polynomials in $\bm{Q}$:
	\begin{equation}\label{eq:cartan-projectors}
		\begin{aligned}
		\bm{\Lambda}_1
		&:= \tfrac{1}{4}(\mathsf{1}_{\mathrm{op}}+\bm{Q})(\mathsf{1}_{\mathrm{op}}+\bm{Q}^2),\\
		\bm{\Lambda}_\pi
		&:= \tfrac{1}{4}(\mathsf{1}_{\mathrm{op}}-\bm{Q})(\mathsf{1}_{\mathrm{op}}+\bm{Q}^2),\\
		\bm{\Lambda}_q
		&:= \tfrac{1}{2}(\mathsf{1}_{\mathrm{op}}-\bm{Q}^2).
		\end{aligned}
	\end{equation}
	They satisfy
	\begin{equation}
		\bm{\Lambda}_\alpha^2=\bm{\Lambda}_\alpha,\qquad
		\bm{\Lambda}_\alpha\bm{\Lambda}_\beta=0\;(\alpha\neq\beta),\qquad
		\bm{\Lambda}_1+\bm{\Lambda}_\pi+\bm{\Lambda}_q=\mathsf{1}_{\mathrm{op}},
	\end{equation}
	with $\alpha,\beta\in\{1,\pi,q\}$. These projectors resolve the alignment operator, as distinguished from the sequence decomposition in Remark~\ref{rem:eigen-vs-modal}. {By Proposition~\ref{prop:dft} they are the spectral projectors of the discrete Fourier transform: $\bm{\Lambda}_1+\bm{\Lambda}_\pi$ and $\bm{\Lambda}_q$ are the symmetric and antisymmetric parts under the interchange of phases $b$ and $c$, and the further split of the symmetric part by the eigenvalue $\pm1$ of $\bm{Q}$ has no phase-permutation meaning; it is the Fourier eigenstructure proper.}

	This second resolution is signal-independent and therefore acts naturally on operators. For any real linear map $\bm{A}\in\mathrm{End}(\mathbb{R}^6)$, such as a realified impedance or admittance matrix, define the Cartan-Fortescue blocks
	\begin{equation}\label{eq:operator-blocks}
		\bm{A}_{\alpha\beta}
		\;:=\;
		\bm{\Lambda}_\alpha\,\bm{A}\,\bm{\Lambda}_\beta,
		\qquad \alpha,\beta\in\{1,\pi,q\}.
	\end{equation}
	The diagonal blocks describe the action of $\bm{A}$ inside the fixed, half-turn and quadrature sectors; the off-diagonal blocks quantify couplings between those intrinsic sectors of the Fortescue rotation. Thus \eqref{eq:operator-blocks} gives an operator-level diagnostic complementary to the usual sequence-domain blocks $\bm{\Pi}_k\bm{A}\bm{\Pi}_\ell$.

The orthogonal projection of $\bm A$ onto the commutant of $\bm Q$ is its finite-group average
\begin{equation}\label{eq:operator-twirl}
\bm A^{\mathrm{eq}}=\tfrac14\sum_{m=0}^3\bm Q^m\bm A\bm Q^{-m}.
\end{equation}
All operator norms below are Frobenius norms. The residual includes off-diagonal Cartan blocks and, for a general real operator, any part of the quadrature diagonal block that fails to commute with $\mathcal J$. Vanishing off-diagonal blocks alone is therefore not sufficient for $\bm Q$-equivariance.

The phase-permutation content can be separated from the remaining Fourier-frame content. Put
$\bm A^{\mathrm{sw}}=\tfrac12(\bm A+\bm Q^2\bm A\bm Q^{-2})$.
Since the commutant of $\bm Q$ is contained in that of $\bm Q^2$, these are nested orthogonal projections and
\begin{equation}\label{eq:nested-residual}
\|\bm A-\bm A^{\mathrm{eq}}\|_F^2
=\|\bm A-\bm A^{\mathrm{sw}}\|_F^2
+\|\bm A^{\mathrm{sw}}-\bm A^{\mathrm{eq}}\|_F^2.
\end{equation}
The first term measures failure of the $b,c$ interchange symmetry. The second measures the additional Fourier-frame restriction. For complex-linear operators it consists precisely of the $(1,\pi)$ and $(\pi,1)$ blocks; for general real operators it also includes quadrature anisotropy. This separates a phase-symmetry test from a convention-dependent alignment diagnostic.

\paragraph{Verification.} The structural script directly evaluates the 64-term seeded sum in Supplementary Section~S1 and checks its coefficient and normalisation, frame alignment, \eqref{eq:rotor-diag}, the fractional path, both commutant dimensions and \eqref{eq:nested-residual}.

\subsubsection{Complex structure, conjugate coupling and approximation}
\label{subsubsec:J-equivariance}

The internal complex structure supplies a second operator symmetry. A real-linear $\bm A$ commutes with $\mathcal J$ exactly when it is the realification of a complex-linear map. Its unique splitting is
\begin{equation}\label{eq:J-split}
\bm A=\bm A_++\bm A_-,\qquad
\bm A_\pm=\tfrac12(\bm A\mp\mathcal J\bm A\mathcal J),\qquad
\mathcal J\bm A_\pm=\pm\bm A_\pm\mathcal J.
\end{equation}
In complex coordinates this reads
\begin{equation}\label{eq:Y1Y2}
\phasor I=\bm Y_1\phasor V+\bm Y_2\overline{\phasor V},
\end{equation}
with $\bm A_+=\mathfrak R(\bm Y_1)$ and $\bm A_-=\mathfrak R(\bm Y_2)\bm K$, where $\bm K\iota(z)=\iota(\bar z)$. This is the established widely linear decomposition~\cite{picinbono1995,schreier2010}; related complex and conjugate transfer operators are used in three-phase dynamics~\cite{harnefors2007complex}. Statistical impropriety concerns signal covariances and is not synonymous with a nonzero $\bm A_-$: conjugation preserves propriety of a proper input signal. The geometric issue here is the interaction of \eqref{eq:J-split} with cyclic symmetry.

\begin{theorem}[Splitting of cyclically symmetric operators]\label{thm:splitting}
If $[\bm A,\bm S]=0$, then $\bm A_+$ and $\bm A_-$ commute with $\bm S$ separately, and
\begin{equation}\label{eq:Aplus-diagonal}
\bm A_+=\sum_{k\in\{0,+,-\}}(\alpha_k\bm\Pi_k+\beta_k\mathcal J\bm\Pi_k),
\qquad \alpha_k,\beta_k\in\mathbb R.
\end{equation}
The conjugate part obeys
\begin{equation}\label{eq:Aminus-exchange}
\bm A_-\bm B_0\subseteq\bm B_0,\qquad
\bm A_-\bm B_+\subseteq\bm B_-,\qquad
\bm A_-\bm B_-\subseteq\bm B_+.
\end{equation}
The real commutant of $\bm S$ has dimension 12, six for each part.
\end{theorem}
\begin{proof}
The relations $[\bm S,\bm A]=[\bm S,\mathcal J]=0$ imply $[\bm S,\bm A_\pm]=0$. A complex-linear map commuting with $S$ preserves each of its three simple complex eigenlines, giving \eqref{eq:Aplus-diagonal}. For the conjugate-linear part, $Sz=\lambda z$ implies
$S A_-z=A_-Sz=\bar\lambda A_-z$, which gives \eqref{eq:Aminus-exchange}. On each allowed source--target pair the map is specified by one complex scalar. There are three such pairs for each part, and every such choice commutes with $S$.
\end{proof}

\begin{corollary}\label{cor:coupling}
For $[\bm A,\bm S]=0$, the positive and negative diagonal sequence blocks are complex-linear:
$\bm\Pi_k\bm A\bm\Pi_k=\bm\Pi_k\bm A_+\bm\Pi_k$ for $k=+,-$.
The only inter-sequence blocks are $\bm\Pi_\pm\bm A_-\bm\Pi_\mp$, and are conjugate-linear. The zero-sequence diagonal block may contain both parts.
\end{corollary}
The zero-sequence exception is necessary: $\bm K$ commutes with $\bm S$ but acts by conjugation on $\bm B_0$. More generally, the same argument for the $N$-phase cyclic shift gives the selection rule $k\mapsto-k\pmod N$ for conjugate blocks; the self-conjugate sectors are $k=0$ and, for even $N$, $k=N/2$.

\begin{prop}[Optimal complex-linear approximation]\label{prop:best-approx}
Let $\mathcal J$ be an orthogonal complex structure on a real Euclidean space, including the present $\mathbb R^6$. The splitting \eqref{eq:J-split} is Frobenius-orthogonal. For $\bm A\neq0$, define
\begin{equation}\label{eq:J-twirl}
\bm A^{\mathrm{eq}_{\mathcal J}}=\tfrac14\sum_{m=0}^3\mathcal J^m\bm A\mathcal J^{-m}=\bm A_+,
\qquad \rho_{\mathcal J}(\bm A)=\frac{\|\bm A_-\|_F}{\|\bm A\|_F}.
\end{equation}
Then
\begin{equation}\label{eq:best-complex-approx}
\rho_{\mathcal J}(\bm A)
=\frac{\min_{[\bm B,\mathcal J]=0}\|\bm A-\bm B\|_F}{\|\bm A\|_F}
=\frac{\|[\bm A,\mathcal J]\|_F}{2\|\bm A\|_F}\in[0,1],
\end{equation}
with unique minimiser $\bm B=\bm A_+$. Set $\rho_{\mathcal J}(0)=0$ by convention.
\end{prop}
\begin{proof}
The map $\mathcal T(\bm A)=-\mathcal J\bm A\mathcal J$ is an orthogonal involution for the Frobenius inner product, so its eigenspaces are orthogonal and $(I\pm\mathcal T)/2$ are their orthogonal projectors. For $[\bm B,\mathcal J]=0$,
$\|\bm A-\bm B\|_F^2=\|\bm A_-\|_F^2+\|\bm A_+-\bm B\|_F^2$.
Finally $[\bm A,\mathcal J]=2\bm A_-\mathcal J$ and $\mathcal J$ is orthogonal.
\end{proof}
The index is unchanged by an orthogonal coordinate change applied to both $\bm A$ and $\mathcal J$. It measures an operator approximation error for the specified metric and complex structure, not passivity or stability. Ordinary complex-linear admittances, including non-passive ones, give zero. Moreover, $\bm\Pi_k\bm A_\pm\bm\Pi_\ell$ resolves the linearity type of each sequence block; its norm alone does not.

\paragraph{Frequency interpretation.}
Equation~\eqref{eq:Y1Y2} is algebraic and does not assign frequencies. Mirror-frequency coupling requires a rotating-frame or lifted-frequency model: the modified sequence impedances of~\cite{rygg2016modified,bakhshizadeh2016} explicitly connect different frequencies. A conventional complex admittance matrix always realifies to a $\mathcal J$-linear operator, so its realification alone cannot reveal omitted frequency channels. Section~\ref{sec:ex-mirror} derives a frequency-pair representation from a synchronous-frame model and applies \eqref{eq:best-complex-approx} to its specified complex structure. This keeps the cyclic selection rule distinct from the physical frequency assignment.

	\subsection{Equivalence to Fortescue's Method}\label{sec:fortescue-equiv}

	The classical Fortescue transformation expresses the three sequence phasors as the coordinate map
	\begin{equation}\label{eq:fortescue-matrix}
		\begin{bmatrix} \phasor{V}_0 \\ \phasor{V}_+ \\ \phasor{V}_- \end{bmatrix}
		=
		\frac{1}{3}
		\begin{bmatrix}
			1 & 1 & 1 \\
			1 & \phasor{a} & \phasor{a}^2 \\
			1 & \phasor{a}^2 & \phasor{a}
		\end{bmatrix}
		\begin{bmatrix} \phasor{V}_a \\ \phasor{V}_b \\ \phasor{V}_c \end{bmatrix},	
	\end{equation}
	With the orthonormal convention of Section~\ref{sec:basis-vectors}, the two descriptions are related by
	\begin{equation}\label{eq:fortescue-coordinate-scaling}
		V_k^c = \sqrt{3}\,\mathrm{Re}\,\phasor{V}_k,
		\qquad
		V_k^s = \sqrt{3}\,\mathrm{Im}\,\phasor{V}_k,
		\qquad k\in\{0,+,-\}.
	\end{equation}
	The factor $\sqrt{3}$ is not a discrepancy: it is the consequence of using unit modal basis vectors in $\mathbb{R}^6$, whereas the engineering Fortescue matrix \eqref{eq:fortescue-matrix} uses the amplitude-invariant normalisation.

	The idempotent objects behind \eqref{eq:fortescue-matrix} are not the coordinate matrix itself, but the spectral projectors of the cyclic shift. In $\mathbb{C}^3$, with the unit eigenvectors $\vec{v}_k$ of \eqref{eq:cyclic-eigenvectors}, define
	\begin{equation}\label{eq:complex-spectral-projectors}
		\bm{P}_k^{\mathbb{C}} := v_k v_k^{*},
		\qquad
		k\in\{0,+,-\}.
	\end{equation}
	They satisfy
	\begin{equation}
		(\bm{P}_k^{\mathbb{C}})^2=\bm{P}_k^{\mathbb{C}},
		\qquad
		\bm{P}_k^{\mathbb{C}}\bm{P}_\ell^{\mathbb{C}}=0\;(k\neq\ell),
		\qquad
		\bm{P}_0^{\mathbb{C}}+\bm{P}_+^{\mathbb{C}}+\bm{P}_-^{\mathbb{C}}=\bm{I}_3,
	\end{equation}
	and resolve the cyclic-shift operator as
	\begin{equation}
		\bm{S}
		=
		\bm{P}_0^{\mathbb{C}}
		+\phasor{a}\,\bm{P}_+^{\mathbb{C}}
		+\phasor{a}^2\,\bm{P}_-^{\mathbb{C}},
	\end{equation}
	up to the chosen ordering convention for positive and negative sequence. The real modal projectors are precisely the images of these complex projectors under the embedding $\iota$:
	\begin{equation}\label{eq:real-complex-projector-equiv}
		\bm{\Pi}_k\,\iota(z)
		=
		\iota\!\left(\bm{P}_k^{\mathbb{C}}z\right),
		\qquad z=(\phasor{V}_a,\phasor{V}_b,\phasor{V}_c)^\top\in\mathbb{C}^3.
	\end{equation}
	Thus the geometric decomposition and the classical Fortescue calculation are the same spectral resolution expressed in two different coordinate languages.

	The rotor formulation gives the same equivalence in frame language. Since $\bm{R}\bm{\sigma}_i\widetilde{\bm{R}}=\bm{u}_i$, the inverse sandwich
	\begin{equation}\label{eq:rotor-modal-coordinates}
		\bm{v}_{\mathrm{mod}}
		:=
		\widetilde{\bm{R}}\,\bm{v}\,\bm{R}
	\end{equation}
	writes the same vector in the canonical pair-of-axes frame:
	\begin{equation}
		\bm{v}_{\mathrm{mod}}
		=
		V_0^c\bm{\sigma}_1+V_0^s\bm{\sigma}_2
		+V_+^c\bm{\sigma}_3+V_+^s\bm{\sigma}_4
		+V_-^c\bm{\sigma}_5+V_-^s\bm{\sigma}_6.
	\end{equation}
	Equations \eqref{eq:fortescue-coordinate-scaling} and \eqref{eq:rotor-modal-coordinates} close the triangle between the classical Fortescue coordinates, the real modal projectors and the frame-alignment rotor. The distinction from the alignment's eigenplanes remains that of Remark~\ref{rem:eigen-vs-modal}.

	\begin{remark}[Sparse $\mathsf{P}/\mathsf{N}$ plane-transfer form]\label{rem:PN-plane-transfer}
	The analytic lift also exposes a sparse realisation of the real Fortescue transform. In the phase-grouped reordering of the canonical basis, one can define two families of rank-two partial isometries: an orientation-preserving plane transfer $\mathsf{P}$, which copies an oriented coordinate plane onto another, and an orientation-reversing transfer $\mathsf{N}$, which copies it with the orientation reversed. The real Fortescue stage followed by a phasor Clarke rotation of the output coordinates then factors \emph{exactly} into one unit-weight $\mathsf{P}$ transfer plus four $\mathsf{P}/\mathsf{N}$ transfers of weight $1/\sqrt{2}$, whose coefficients reproduce the real Fourier entries associated with the complex factors $\phasor{a}$ and $\phasor{a}^2$ while the composite remains $\mathcal{J}$-linear and orthogonal. {Some transfer planes are not $\mathcal J$-invariant, so an individual $\mathsf N$ must not be identified with the conjugate part in \eqref{eq:J-split}. The original invariant sequence projectors are recovered by undoing the output rotation before coordinate-pair selection.} Thus the factorisation is a coordinate-level implementation of the modal transform, complementary to the intrinsic rotor description of Section~\ref{sec:rotor-method}. The complete construction --- the dyadic forms of $\mathsf{P}$ and $\mathsf{N}$, their partial-isometry identities and the explicit sparse factorisation --- is given in the Supplementary Material (Section~S2).
    \end{remark}

	\subsection{Norm and Parseval identity}
	\label{sec:parseval}

	The modal projectors form an orthogonal resolution of the identity, hence the signal norm decomposes additively as
	\begin{equation}\label{eq:parseval}
		\|\bm{v}\|^2
		=
		\|\bm{v}_0\|^2+\|\bm{v}_+\|^2+\|\bm{v}_-\|^2
		=
		\sum_{k\in\{0,+,-\}}\bigl\|\bm{\Pi}_k\bm{v}\bigr\|^2.
	\end{equation}
	Each modal summand has the local two-dimensional form
	\begin{equation}\label{eq:modal-local-norm}
		\|\bm{v}_k\|^2
		=
		(V_k^{c})^2+(V_k^{s})^2
		=
		3\,|\phasor{V}_k|^2,
	\end{equation}
	where the last equality uses the scaling relation \eqref{eq:fortescue-coordinate-scaling}. Therefore the real Parseval identity is the embedded version of the classical sequence-energy identity
	\begin{equation}\label{eq:fortescue-parseval}
		\|\bm{v}\|^2
		=
		3\left(|\phasor{V}_0|^2+|\phasor{V}_+|^2+|\phasor{V}_-|^2\right).
	\end{equation}
	In analogy with the multivector norm recently developed by Shirokov~\cite{shirokov2025rank} for complexified geometric algebras, the present real norm decomposes over the modal idempotents without introducing complex coordinates. %

	The frame-alignment rotor preserves the same norm:
	\begin{equation}
		\|\bm{v}\|^2
		=
		\bigl\|\widetilde{\bm{R}}\,\bm{v}\,\bm{R}\bigr\|^2.
	\end{equation}
	This is the rotor version of Parseval's identity: after the inverse sandwich \eqref{eq:rotor-modal-coordinates}, the three sequence energies are simply the squared norms of the three canonical coordinate pairs. Independently, the eigenblade projectors of Section~\ref{subsubsec:fortescue-operator-blocks} give a second orthogonal norm resolution,
	\begin{equation}\label{eq:cartan-parseval}
		\|\bm{v}\|^2
		=
		\|\bm{\Lambda}_1\bm{v}\|^2
		+\|\bm{\Lambda}_\pi\bm{v}\|^2
		+\|\bm{\Lambda}_q\bm{v}\|^2.
	\end{equation}
	Equation~\eqref{eq:cartan-parseval} should not be read as a zero/positive/negative sequence energy balance. It decomposes the same signal with respect to the fixed, half-turn and quadrature eigenblades of the Fortescue alignment rotor, whereas \eqref{eq:parseval} decomposes the signal with respect to the modal sequence planes.

	\subsubsection{Sequence unbalance as an intrinsic ratio.} For nonzero positive sequence, a direct consequence of \eqref{eq:parseval} is that the ratios
	\begin{equation}\label{eq:vuf-like}
		\eta_- \;:=\; \frac{\|\bm{v}_-\|^2}{\|\bm{v}_+\|^2}
		=
		\frac{|\phasor{V}_-|^2}{|\phasor{V}_+|^2},
		\qquad
		\eta_0 \;:=\; \frac{\|\bm{v}_0\|^2}{\|\bm{v}_+\|^2}
		=
		\frac{|\phasor{V}_0|^2}{|\phasor{V}_+|^2}
	\end{equation}
	provide intrinsic, basis-invariant measures of negative-to-positive and zero-to-positive sequence imbalance, derived directly from the modal spectral decomposition. The standard Voltage Unbalance Factor used in power engineering corresponds to $\sqrt{\eta_-}$; zero-sequence content is reported separately by $\eta_0$.

	\section{Extension to multi-harmonic systems}\label{sec:multiharm}

	The single-frequency framework developed in Sections~\ref{sec:basis-vectors}--\ref{sec:parseval} extends naturally to signals containing multiple harmonic components. The construction is straightforward: each harmonic of order $n$ contributes its own copy of $\mathbb{R}^6$ to the total signal space, and the modal decomposition acts independently on each copy. The resulting framework unifies harmonic and unbalance analysis within a single geometric structure.

		\subsection{Unified representation in \texorpdfstring{$\mathbb{R}^{6h}$}{R6h}}\label{sec:Rh}

	Consider a three-phase signal with $h$ harmonics at angular frequencies $\omega, 2\omega, \dots, h\omega$. For each harmonic $n\in\{1,\dots,h\}$, the three phasors $\phasor{V}_{a,n}=V_{a,n}\angle\theta_{a,n}$, $\phasor{V}_{b,n}=V_{b,n}\angle\theta_{b,n}$, $\phasor{V}_{c,n}=V_{c,n}\angle\theta_{c,n}$ admit the same six-dimensional representation introduced in Section~\ref{sec:basis-vectors}: a vector $\bm{v}^{(n)} \in \mathbb{R}^6$ with one cosine/sine pair per phase. The full multi-harmonic signal is the vector
	\begin{equation}\label{eq:multi-rep}
		\bm{v} \;=\; \bm{v}^{(1)} \oplus \bm{v}^{(2)} \oplus \cdots \oplus \bm{v}^{(h)} \;\in\; \mathbb{R}^{6h},
	\end{equation}
	with explicit structure
	\begin{equation}
\begin{aligned}
\bm v=\Bigl[&\underbrace{v_{a,1}^{c},v_{a,1}^{s},v_{b,1}^{c},v_{b,1}^{s},v_{c,1}^{c},v_{c,1}^{s}}_{\text{fundamental block}},\\
&\underbrace{v_{a,2}^{c},v_{a,2}^{s},\dots,v_{c,2}^{s}}_{\text{second harmonic}},\ \dots,\
\underbrace{v_{a,h}^{c},v_{a,h}^{s},\dots,v_{c,h}^{s}}_{h\text{-th harmonic}}\Bigr]^{\!\top}.
\end{aligned}
	\end{equation}
	The ambient space $\mathbb{R}^{6h}$ is the orthogonal direct sum of $h$ copies of $\mathbb{R}^6$, one per harmonic. The underlying geometric algebra is $\Cl_{6h,0}(\mathbb{R})$, although only the modal-projection structure (a linear operation) is needed for the present discussion; no full multivector expansion is required. Any selected set of harmonic orders can be used instead of consecutive orders, with one block per selected frequency.

	\subsection{Modal projectors per harmonic}\label{sec:projectors-multi}

	For each harmonic $n$, the three modal planes $\bm{B}_0^{(n)}, \bm{B}_{+}^{(n)}, \bm{B}_{-}^{(n)}$ live inside the $n$-th block $\mathbb{R}^6_{(n)}\subset\mathbb{R}^{6h}$, and each has its own orthogonal idempotent projector
	\begin{equation}\label{eq:Pi-nk}
		\bm{\Pi}_k^{(n)} \;\in\; \mathrm{End}(\mathbb{R}^{6h}), \qquad k\in\{0,+,-\},\quad n\in\{1,\dots,h\}.
	\end{equation}
	The projector $\bm{\Pi}_k^{(n)}$ acts as the single-harmonic projector $\bm{\Pi}_k$ of Section~\ref{subsec:spectral} on the $n$-th block of coordinates and as zero on every other block; equivalently, it is the matrix-direct-sum
	\begin{equation}
		\bm{\Pi}_k^{(n)} \;=\; \bm{0} \oplus \cdots \oplus \underbrace{\bm{\Pi}_k}_{n\text{-th block}} \oplus \cdots \oplus \bm{0}.
	\end{equation}

	The total family $\{\bm{\Pi}_k^{(n)}\;:\; n=1,\dots,h,\; k\in\{0,+,-\}\}$ consists of $3h$ projectors, satisfying the complete-orthogonal-system identities
	\begin{equation}\label{eq:full-system}
		(\bm{\Pi}_k^{(n)})^2 = \bm{\Pi}_k^{(n)},\quad
		\bm{\Pi}_k^{(n)}\bm{\Pi}_\ell^{(m)} = \delta_{k\ell}\delta_{nm}\bm{\Pi}_k^{(n)},\quad
		\sum_{n=1}^{h}\sum_{k\in\{0,+,-\}}\bm{\Pi}_k^{(n)} = \mathsf{1}_{\mathrm{op}}.
	\end{equation}
	The product of Kronecker deltas $\delta_{k\ell}\delta_{nm}$ encodes the simultaneous orthogonality of distinct modal planes within each harmonic block and of distinct harmonic blocks across the full space. Consequently, $\mathbb{R}^{6h}$ decomposes as the orthogonal direct sum of $3h$ two-dimensional modal subspaces:
	\begin{equation}
		\mathbb{R}^{6h} \;=\; \bigoplus_{n=1}^{h}\bigoplus_{k\in\{0,+,-\}}\operatorname{im}\bm{\Pi}_k^{(n)}.
	\end{equation}

	\subsubsection{Connection with the classical harmonic sequence rule.} For balanced waveforms related by one-third-period phase delays, the familiar classification of harmonic orders is recovered as a special case of the projector system above~\cite{chicco2019}. For harmonic order $n$,
	\begin{equation}\label{eq:harmonic-sequence-rule}
	\begin{gathered}
		n\equiv 1 \pmod{3} \Rightarrow \bm{\Pi}_+^{(n)},\qquad
		n\equiv 2 \pmod{3} \Rightarrow \bm{\Pi}_-^{(n)},\\
		n\equiv 0 \pmod{3} \Rightarrow \bm{\Pi}_0^{(n)}.
	\end{gathered}
	\end{equation}
	Thus positive-sequence harmonics, negative-sequence harmonics and triplen zero-sequence harmonics correspond to the cases in which only the compatible modal projector has non-zero output in the $n$-th block. Under unbalance this exclusivity need not hold: all three projectors may carry non-zero energy at the same harmonic order. This caveat is not academic. The identification of triplen harmonics with zero sequence is widely quoted as if it were unconditional, whereas in unbalanced networks the triplen orders carry positive- and negative-sequence content as well, and treating the rule as exact leads to incorrect conclusions about neutral currents and compensation~\cite{chicco2011triplen}. The three-projector resolution therefore remains necessary at every harmonic order.

	\subsection{Selective extraction and reference generation}\label{sec:selective}

	The orthogonality of the $3h$ modal subspaces enables the extraction of any sequence--harmonic combination by a single projection. For instance, the forward sequence of the fundamental is
	\begin{equation}\label{eq:forward-fund}
		\bm{v}_{+}^{(1)} \;=\; \bm{\Pi}_{+}^{(1)}\,\bm{v},
	\end{equation}
	a single linear operation that simultaneously rejects (i)~all harmonics other than the fundamental and (ii)~the backward and common sequences within the fundamental. The output $\bm{v}_{+}^{(1)}$ represents a sinusoidal, balanced three-phase signal at the fundamental frequency. 

	\subsubsection{Active filter reference.} A direct application is the generation of compensation references for shunt active power filters~\cite{akagi}. Three-phase currents admit the same embedding into $\mathbb{R}^{6h}$ as voltages, with $\iota$ acting phase-by-phase on the harmonic current phasors; we write $\bm{i}$ rather than $\bm{v}$ in this paragraph to keep the engineering convention. The non-fundamental, non-positive-sequence content of a measured three-phase current $\bm{i}\in\mathbb{R}^{6h}$ is the residual
	\begin{equation}\label{eq:filter-ref}
		\bm{i}_{\text{ref}} \;=\; \bm{i} \;-\; \bm{\Pi}_{+}^{(1)}\,\bm{i} \;=\; \sum_{(k,n)\neq(+,1)}\bm{\Pi}_k^{(n)}\,\bm{i}.
	\end{equation}
	The reference $\bm{i}_{\text{ref}}$ contains, by construction, all retained higher harmonics and all zero- and negative-sequence content simultaneously, using a single projection of the already assembled harmonic-coordinate vector. This does not replace harmonic or phasor estimation; rather, once those coordinates are available, no additional sequence-by-sequence transform is required because the modal decomposition is already aligned with both harmonic order and sequence type. This reference does not impose a power-factor objective on the retained positive-sequence fundamental. Frequency transfer by a device requires a separate operator with blocks between frequency channels.

	\subsubsection{Parseval identity in \texorpdfstring{$\mathbb{R}^{6h}$}{R6h}.} The Parseval identity \eqref{eq:parseval} extends to the multi-harmonic setting as
	\begin{equation}\label{eq:parseval-h}
		\|\bm{v}\|^2 \;=\; \sum_{n=1}^{h}\sum_{k\in\{0,+,-\}}\bigl\|\bm{\Pi}_k^{(n)}\bm{v}\bigr\|^2,
	\end{equation}
	which provides a single basis-invariant measure of the total signal energy decomposed jointly across harmonic order and sequence type.

	\section{Worked examples and verification}\label{sec:examples}

	This section gives five reproducible examples. The first checks that the classical Fortescue phasors, the real modal projectors and the frame-alignment rotor give the same coordinates. The second illustrates selective extraction in the multi-harmonic space $\mathbb{R}^{6h}$. The third applies the two projector systems to an admittance operator, showing that the modal sequence blocks and the Cartan/eigenblade blocks diagnose different symmetries of the same matrix. The fourth compares complex-linear and conjugate-linear couplings with equal sequence-block norms, then derives a frequency-pair response from synchronous-frame dynamics. The fifth uses a public nonlinear machine model to assess the consequence of the projection for a compensation decision. A separate structural script checks the propositions and the seeded rotor sum; numerical checks supplement their proofs.

	\subsection{Signal-level example: unbalanced three-phase set}\label{sec:ex-signal}

	Consider the unbalanced phasor set
	\begin{equation}\label{eq:unbalanced-input}
		\phasor{V}_a = 230\angle 0^\circ,\qquad
		\phasor{V}_b = 210\angle{-110^\circ},\qquad
		\phasor{V}_c = 240\angle 130^\circ .
	\end{equation}
	After the embedding $\iota:\mathbb{C}^3\to\mathbb{R}^6$, the modal coordinates are obtained either by inner products with the modal frame or by the inverse rotor sandwich $\widetilde{\bm{R}}\bm{v}\bm{R}$. Table~\ref{tab:signal-example} reports the classical Fortescue phasor magnitude $|\phasor{V}_k|$, its phase, the corresponding real modal norm $\|\bm{v}_k\|=\sqrt{3}|\phasor{V}_k|$, and the energy fraction $\|\bm{v}_k\|^2/\|\bm{v}\|^2$.

	\begin{table}[htbp]
	\centering
	\renewcommand{\arraystretch}{1.2}
	\caption{Signal-level decomposition of the unbalanced phasor set \eqref{eq:unbalanced-input}.}
	\label{tab:signal-example}
	\begin{tabular}{lrrrr}
	\toprule
	\textbf{Sequence} & $\left|\phasor{V}_k\right|$ & \textbf{Phase} & $\|\bm{v}_k\|$ & \textbf{Energy share} \\
	\midrule
	Zero $(0)$ & $4.680$ & $-73.84^\circ$ & $8.106$ & $0.0425\%$ \\
	Positive $(+)$ & $225.895$ & $\phantom{-}6.62^\circ$ & $391.261$ & $99.0201\%$ \\
	Negative $(-)$ & $21.979$ & $-78.69^\circ$ & $38.069$ & $0.9374\%$ \\
	\bottomrule
	\end{tabular}
	\end{table}

	The coordinate-level agreement between the real modal projection and the classical complex Fortescue calculation is below $7.4\times10^{-14}$ for all three sequence phasors. The rotor calculation gives the same modal coordinate vector, with
	\[
		\max_i\left|
		\bigl(\widetilde{\bm{R}}\bm{v}\bm{R}\bigr)_i
		-
		(\bm{U}^\top\bm{v})_i
		\right|
		<5.2\times10^{-14},
	\]
	where $\bm{U}$ is the modal-frame matrix. Finally, the Parseval residual
	\[
		\left|\|\bm{v}\|^2-\sum_{k\in\{0,+,-\}}\|\bm{\Pi}_k\bm{v}\|^2\right|
	\]
	is $2.9\times10^{-11}$ in double precision, and the local bivectors satisfy $\bm{B}_0^2=\bm{B}_+^2=\bm{B}_-^2=-1$ numerically to machine precision. This single test therefore verifies the three equivalent descriptions developed in Sections~\ref{subsec:spectral}, \ref{subsubsec:rotor-diag} and~\ref{sec:fortescue-equiv}.

	\subsection{Multi-harmonic example: selective extraction}\label{sec:ex-multiharm}

	For a multi-harmonic signal, each harmonic block has its own copy of the sequence projector triple. Table~\ref{tab:multiharm-example} defines a synthetic three-phase signal containing a dominant fundamental positive sequence, a small fundamental unbalance, a third-harmonic zero sequence, a fifth-harmonic negative sequence and a seventh-harmonic positive sequence. The phase phasors are generated by the inverse Fortescue map and then embedded in $\mathbb{R}^{24}$.

	\begin{table}[htbp]
	\centering
	\renewcommand{\arraystretch}{1.2}
	\caption{Sequence content of the multi-harmonic example.}
	\label{tab:multiharm-example}
	\begin{tabular}{rrrr}
	\toprule
	\textbf{Harmonic} & $\left|\phasor{V}_0\right|$ & $\left|\phasor{V}_+\right|$ & $\left|\phasor{V}_-\right|$ \\
	\midrule
	$\#1$ & $2.0$ & $230.0$ & $6.0$ \\
	$\#3$ & $18.0$ & $0.0$ & $0.0$ \\
	$\#5$ & $0.0$ & $0.0$ & $13.0$ \\
	$\#7$ & $0.0$ & $10.0$ & $0.0$ \\
	\bottomrule
	\end{tabular}
	\end{table}

	The projector $\bm{\Pi}_+^{(1)}$ extracts the desired fundamental positive-sequence component as one linear operation on the stacked vector. In this example
	\[
		\|\bm{v}\|^2 = 160599,\qquad
		\|\bm{\Pi}_+^{(1)}\bm{v}\|^2 = 158700,\qquad
		\|\bm{v}-\bm{\Pi}_+^{(1)}\bm{v}\|^2 = 1899,
	\]
	with a Parseval discrepancy below $9\times10^{-11}$. Thus $\bm{\Pi}_+^{(1)}$ retains $98.8176\%$ of the total energy and the residual contains, by construction, every undesired component: fundamental zero and negative sequence, third-harmonic zero sequence, fifth-harmonic negative sequence and seventh-harmonic positive sequence. This is the practical content of \eqref{eq:filter-ref}: once the harmonic phasors have been assembled, harmonic selection and sequence selection are not separate algorithms but different idempotents in one direct-sum algebra.

	\subsection{Operator example: modal blocks versus eigenblade blocks}\label{sec:ex-operator}

	The signal examples above do not use the full force of the eigenblade decomposition, because signal energies are naturally organised by the modal projectors $\bm{\Pi}_0,\bm{\Pi}_+,\bm{\Pi}_-$. The new decomposition becomes more informative for operators. Let
	\[
		\bm{Y}=
		\begin{bmatrix}
		y_a+y_{ab}+y_{ca} & -y_{ab} & -y_{ca}\\
		-y_{ab} & y_b+y_{ab}+y_{bc} & -y_{bc}\\
		-y_{ca} & -y_{bc} & y_c+y_{bc}+y_{ca}
		\end{bmatrix}
	\]
	be a simple three-phase nodal admittance, realified as an operator on $\mathbb{R}^6$. The balanced case uses $y_a=y_b=y_c=0.85-0.18j$ and $y_{ab}=y_{bc}=y_{ca}=0.12-0.04j$. The asymmetric case uses $(y_a,y_b,y_c)=(0.93-0.17j,0.78-0.22j,0.86-0.15j)$ and $(y_{ab},y_{bc},y_{ca})=(0.13-0.035j,0.10-0.050j,0.145-0.030j)$. Table~\ref{tab:operator-ratios} compares the modal off-diagonal ratio and the relative residuals $\|\bm A-\bm A^{\mathrm{sw}}\|_F/\|\bm A\|_F$ and $\|\bm A-\bm A^{\mathrm{eq}}\|_F/\|\bm A\|_F$. For these complex-linear examples the Fourier residual equals the Cartan off-diagonal ratio.

	\begin{table}[htbp]
	\centering
	\renewcommand{\arraystretch}{1.2}
	\caption{Complementary operator diagnostics from modal and Cartan blocks.}
	\label{tab:operator-ratios}
	\scriptsize
	\setlength{\tabcolsep}{4pt}
	\begin{tabular}{lrrrr}
	\toprule
	\textbf{Operator} & \textbf{Modal off-diag.} & \textbf{Cartan off-diag.} & \textbf{Swap residual} & \textbf{Fourier residual} \\
	\midrule
	Balanced $\bm{Y}_{\mathrm{bal}}$ & $0.000000$ & $0.111469$ & $0.000000$ & $0.111469$ \\
	Asymmetric $\bm{Y}_{\mathrm{asym}}$ & $0.079168$ & $0.158793$ & $0.043521$ & $0.158793$ \\
	Equivariant part $\bm{Y}_{\mathrm{asym}}^{\mathrm{eq}}$ & $0.061061$ & $0.000000$ & $0.000000$ & $0.000000$ \\
	\bottomrule
	\end{tabular}
	\end{table}

	{The balanced reciprocal admittance respects every phase interchange, so its swap residual vanishes. Its $(1,\pi)$ and $(\pi,1)$ Cartan blocks nevertheless each have Frobenius norm $0.2191$: this contribution belongs entirely to the second term in \eqref{eq:nested-residual}. The asymmetric case breaks the $b,c$ symmetry and also retains coupling within the swap-symmetric sector: the $(1,\pi)$, $(1,q)$ and $(\pi,q)$ blocks have Frobenius norms $0.3054$, $0.0266$ and $0.0829$, respectively. Averaging over $\bm Q$ removes both contributions but does not enforce cyclic balance, as its remaining modal ratio shows. The nested residual therefore identifies which portion of the Fourier-frame diagnostic is attributable to an actual phase interchange.}

	\subsection{Conjugate coupling: algebraic and frequency-pair examples}\label{sec:ex-mirror}

\paragraph{Algebraic benchmark.}
In modal coordinates, set every diagonal $2\times2$ block to the realification of $y_0=0.8-0.3j$. Use either zero off-diagonal blocks, complex-linear $(+,-)$ and $(-,+)$ blocks $z\mapsto cz$, or conjugate-linear blocks $z\mapsto c\bar z$, with $c=0.22/\sqrt2$. Transforming back with $\bm Q$ gives the three real operators in Table~\ref{tab:mirror-example}.
\begin{table}[htbp]
\centering
\caption{Equal sequence-block norms can conceal different linearity types. These are algebraic benchmark operators.}
\label{tab:mirror-example}
\begin{tabular}{lrrr}
\toprule
Operator & $\|\bm\Pi_+\bm A\bm\Pi_-\|_F$ & $\rho_{\mathcal J}$ & Fourier residual \\
\midrule
No coupling & $0.0000$ & $0.0000$ & $0.0000$\\
Complex-linear coupling & $0.2200$ & $0.0000$ & $0.0600$\\
Conjugate-linear coupling & $0.2200$ & $0.1470$ & $0.1201$\\
\bottomrule
\end{tabular}
\end{table}
The last two cases have equal sequence-block norms, but their optimal complex-linear approximation errors differ. The conjugate case also commutes with $\bm S$, consistently with Theorem~\ref{thm:splitting}. Recovering its $(+,-)$ block after \eqref{eq:J-split} returns the prescribed $c$. No frequency response is implied until the channels and dynamics are specified.

\paragraph{A synchronous-frame admittance.}
Consider a zero-sequence-free terminal with real $dq$ dynamics
\begin{equation}\label{eq:dq-model}
\tau\dot i_d+i_d=g_d v_d,\qquad \tau\dot i_q+i_q=g_q v_q,
\qquad g_d,g_q,\tau>0.
\end{equation}
This is a minimal first-order admittance model with unequal axis gains, not a full converter controller. Let $v=v_\alpha+jv_\beta$ and $i=i_\alpha+ji_\beta$ be power-invariant Clarke space vectors, and set $z_v=e^{-j\omega_0t}v=v_d+jv_q$, $z_i=e^{-j\omega_0t}i$. Then
\begin{equation}\label{eq:dq-conjugate}
\tau\dot z_i+z_i=g z_v+d\bar z_v,\qquad
g=\frac{g_d+g_q}{2},\quad d=\frac{g_d-g_q}{2}.
\end{equation}
For $\Omega>0$, use the signed space-vector frequencies $\omega_u=\omega_0+\Omega$, $\omega_l=\omega_0-\Omega$ and input
$v(t)=V_u e^{j\omega_ut}+V_l e^{j\omega_lt}$. With $h=(1+j\tau\Omega)^{-1}$, its steady-state output is
\begin{equation}\label{eq:sideband-map}
I_u=h(gV_u+d\overline{V_l}),\qquad
I_l=\bar h(gV_l+d\overline{V_u}).
\end{equation}
Thus an upper-sideband input creates a lower-sideband response: the frequency reflection follows from the rotating-frame dynamics, in the setting underlying modified sequence models~\cite{harnefors2007complex,rygg2016modified}. The labels $u,l$ denote frequency channels, not the same-frequency planes $\bm B_+,\bm B_-$.

Realify \eqref{eq:sideband-map} on the four coordinates of $(V_u,V_l)$, with $J_4=\operatorname{diag}(J_2,J_2)$ and $J_2=\left[\begin{smallmatrix}0&-1\\1&0\end{smallmatrix}\right]$. Proposition~\ref{prop:best-approx} gives
\begin{equation}\label{eq:dq-rho}
\rho_{J_4}=\frac{|d|}{\sqrt{g^2+d^2}}.
\end{equation}
For $g_d=1.0\,\mathrm S$, $g_q=0.6\,\mathrm S$, $\tau=5\,\mathrm{ms}$, $\omega_0/(2\pi)=50\,\mathrm{Hz}$ and $\Omega/(2\pi)=30\,\mathrm{Hz}$, a unit upper-sideband voltage at $80\,\mathrm{Hz}$ gives $|I_u|=0.582182\,\mathrm A$ and $|I_l|=0.145545\,\mathrm A$ at $20\,\mathrm{Hz}$, with $\rho_{J_4}=0.242536$. Direct integration of \eqref{eq:dq-model} and Fourier extraction verify the complex response coefficients. Setting $g_d=g_q$ removes both the mirror response and the residual.

This example also separates coupling from passivity. The storage
$W=\tfrac\tau2(i_d^2/g_d+i_q^2/g_q)$ satisfies
$v_di_d+v_qi_q-\dot W=i_d^2/g_d+i_q^2/g_q\geq0$, although $d\ne0$. Finally, conjugating the lower coordinate makes \eqref{eq:sideband-map} an ordinary augmented complex matrix; the same coordinate change must send $J_4$ to $\operatorname{diag}(J_2,-J_2)$. Keeping the default complex structure instead would change the approximation problem and erase its intended interpretation.

\subsection{Nonlinear machine: consequence for torque-ripple compensation}
\label{sec:ex-machine}

The frequency-pair example above supplies a bridge from the operator splitting to a physical application. Fortescue extraction and the real modal projectors give the same sequence coordinates; neither operation alone makes the device response independent across channels. Here we ask whether replacing a local machine admittance by its optimal complex-linear projection changes the voltage calculated to suppress a torque harmonic.

\paragraph{Machine and frequency channels}
We use the FEM-Parameterized PMSM from the public MathWorks \emph{Three-Phase PMSM Traction Drive} example in MATLAB/Simulink R2024a Update~1~\cite{mathworks2024traction,mathworks2024fempmsm}. The original current-dependent $d/q$ magnetic tables, smooth interpolation and iron-loss settings are retained. The external supply is replaced by ideal three-phase voltage sources and the mechanical speed is imposed. With eight pole pairs, 1500~rpm gives $f_0=200$~Hz; the measured base torque is 152.55~N\,m, approximately 23.96~kW. The fundamental voltage is 113.294~V peak per phase. The case was selected from a disclosed screening of 210 conditions, of which 183 passed the proxy feasibility constraints, followed by three nonlinear-model checks (Supplementary Section~S3). It is a demonstrative operating point, not an estimate of average performance.

\begin{figure}[tbp]
\centering
\includegraphics[width=\linewidth]{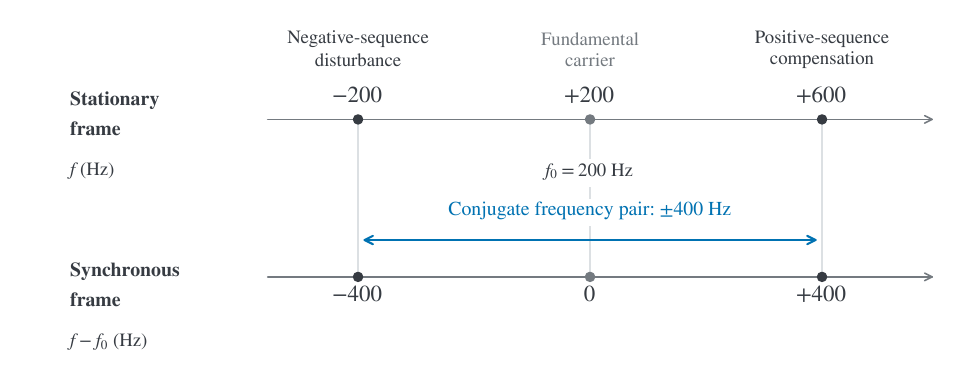}
\caption{Frequency assignment for the machine example. Signed stationary frequencies $(-f_0,3f_0)$ become $(-2f_0,2f_0)$ after subtracting the rotor electrical frequency. The central marker denotes the fundamental carrier. The diagram specifies the channels to which the conjugate splitting is applied; it does not identify them with positive and negative sequence planes at a single frequency.}
\label{fig:machine-channels}
\end{figure}

The disturbance is a 2\% negative-sequence fundamental voltage. In the signed space-vector convention of \eqref{eq:sideband-map}, its channel is $f_l=-f_0=-200$~Hz; compensation is applied at $f_u=3f_0=600$~Hz with positive sequence. Both become opposite offsets $\pm 2f_0$ in the synchronous frame (Fig.~\ref{fig:machine-channels}), allowing an algebraic compensation of the 400~Hz torque harmonic. Positive sequence at the third harmonic is imposed explicitly; the balanced-waveform modulo-three rule does not constrain this independently specified component.

\paragraph{Identification and controlled operator restriction}
Let $\bm x=(\Re V_u,\Im V_u,\Re V_l,\Im V_l)^\top$, and define $\bm y$ similarly from the incremental currents. Eight simulations with signed 0.25~V coordinate probes identify $\bm y\simeq\bm A\bm x$ and $\bm q\simeq\bm G\bm x$, where $\bm q\in\mathbb R^2$ contains the real and imaginary parts of the incremental 400~Hz torque coefficient. Central differences and measured terminal voltages are used. With $J_4=\operatorname{diag}(J_2,J_2)$, Proposition~\ref{prop:best-approx} gives
\begin{equation}\label{eq:machine-projection}
\bm A_+=\tfrac12(\bm A-J_4\bm A J_4),\qquad
\rho_{J_4}=0.21106,\qquad
\bm C=\bm G\bm A^{-1},\quad \bm G_+=\bm C\bm A_+.
\end{equation}
Thus the same local current--torque map $\bm C$ is used in both designs. Writing $\bm G=[\bm G_u\ \bm G_l]$ and $\bm G_+=[\bm G_{+,u}\ \bm G_{+,l}]$, the corrective voltage coordinates are
\begin{equation}\label{eq:machine-compensation}
\bm x_u^{\rm full}=-\bm G_u^{-1}\bm G_l\bm x_l,
\qquad
\bm x_u^{\rm proj}=-\bm G_{+,u}^{-1}\bm G_{+,l}\bm x_l.
\end{equation}
The prescribed disturbance phase is 0.37~rad, distinct from the probe phases. Each voltage is applied once to the original nonlinear machine, without validation-output retuning. The operator acts on four real coordinates of a specified frequency pair, not one same-frequency $\mathbb R^6$ block. Here the subscripts on $\bm A_\pm$ denote linearity type, not sequence; the cyclic-commutation hypothesis of Theorem~\ref{thm:splitting} is not assumed for this test.

\begin{figure}[tbp]
\centering
\includegraphics[width=\linewidth]{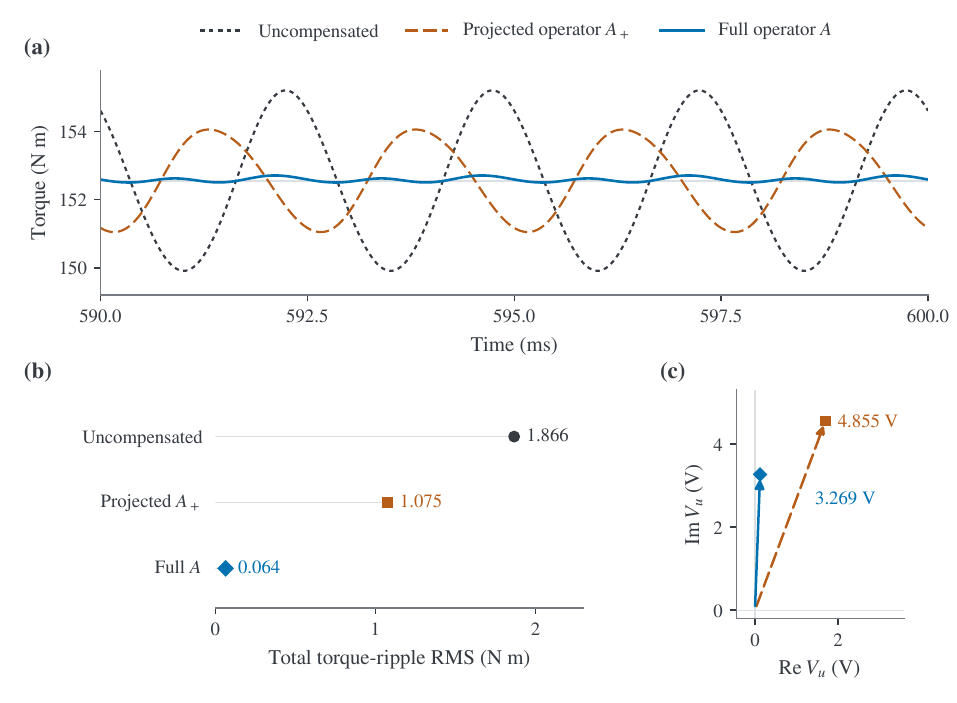}
\caption{Nonlinear-machine validation of the two voltages from \eqref{eq:machine-compensation}. (a) Unfiltered electromagnetic torque over four ripple periods; the displayed interval is a subset of the 0.4--0.6~s evaluation window. (b) RMS of the entire mean-removed torque waveform over that window, including components outside the targeted harmonic. (c) Corrective voltage phasors in the same stationary-channel convention. The full and projected designs use the same current--torque map.}
\label{fig:machine-compensation}
\end{figure}

\paragraph{Nonlinear validation and interpretation}
Figure~\ref{fig:machine-compensation} reports a reduction of total torque-ripple RMS from 1.86632 to 0.06445~N\,m (96.55\%) with the full operator, compared with 1.07525~N\,m (42.39\%) using $\bm A_+$. The full design requires 3.269~V of corrective amplitude, versus 4.855~V for the projection. Its mean torque is 152.590~N\,m, and stator copper losses increase by 0.352\% relative to the uncompensated case. The targeted torque amplitude falls from 2.63869 to 0.04772~N\,m; the residual reflects the nonlinear validation rather than the exact cancellation predicted by the local model. Halving the maximum integration step and tightening the relative tolerance tenfold changes the full-design ripple RMS by $1.90\times10^{-5}$~N\,m. Additional operating points, held-out phases and numerical limits are recorded in Supplementary Section~S3.

This example tests the practical consequence of a specified operator projection. Frobenius optimality does not imply optimal torque compensation, and $\rho_{J_4}$ is not a universal waveform-error percentage. A complete classical $dq$ or augmented complex representation can retain the same coupling~\cite{rygg2016modified}; the comparison is between $\bm A$ and $\bm A_+$, not between different physical theories. The machine and torque maps are identified locally, not derived solely from the geometric theorems. Ideal voltage actuation and imposed speed delimit the example: inverter switching and closed-loop implementation are outside its scope.

\section{Discussion}\label{sec:discussion}

	{Geometrically, the whole construction is a statement about three families of oriented planes in $\mathbb{R}^6$: the phase planes $\bm{I}_A,\bm{I}_B,\bm{I}_C$, where measurements live; the modal planes $\bm{B}_0,\bm{B}_+,\bm{B}_-$, where sequences live; and the Cartan planes $\bm{B}_1,\bm{B}_\pi,\bm{B}_q$, where the rotation that carries the first family onto the second acts. The second family is the image of the first under the Fourier rotor, and the third is fixed by it. }The construction separates four levels that are often compressed into one complex calculation. Table~\ref{tab:interpretation-levels} summarises the role and limitation of each level.

	\begin{table}[htbp]
	\centering
	\renewcommand{\arraystretch}{1.18}
	\caption{Signal, harmonic and operator interpretations of the projector framework.}
	\label{tab:interpretation-levels}
	\scriptsize
	\setlength{\tabcolsep}{3pt}
	\begin{tabular}{>{\raggedright\arraybackslash}p{0.22\linewidth}>{\raggedright\arraybackslash}p{0.15\linewidth}>{\raggedright\arraybackslash}p{0.30\linewidth}>{\raggedright\arraybackslash}p{0.20\linewidth}}
	\toprule
	\textbf{Level} & \textbf{Objects} & \textbf{Meaning} & \textbf{Caution} \\
	\midrule
	Signal decomposition & $\bm{\Pi}_0,\bm{\Pi}_+,\bm{\Pi}_-$ & Real spectral form of Fortescue zero, positive and negative sequence components. & Equivalent to classical Fortescue coordinates. \\
	Multi-harmonic extraction & $\bm{\Pi}_k^{(n)}$ on $\mathbb{R}^{6h}$ & One idempotent selects any harmonic--sequence component after phasors have been estimated. & Does not replace phasor or harmonic estimation. \\
	Operator symmetry & $\bm{\Lambda}_1,\bm{\Lambda}_\pi,\bm{\Lambda}_q$ & Eigenblade blocks of the Fortescue alignment rotor; diagnose couplings between fixed, half-turn and quadrature sectors. {These are the Fourier eigenprojectors; $\bm{\Lambda}_q$ is the antisymmetric part under the interchange of phases $b$ and $c$.} & Not a voltage-unbalance factor or sequence-energy balance. \\
	{Complex structure} & {$\mathcal{J}$, $\bm{A}=\bm{A}_++\bm{A}_-$} & {Complex-linear versus conjugate-linear part of any real three-phase operator; $\rho_{\mathcal J}$ measures optimal complex-linear approximation error. Cyclic symmetry allows conjugate exchange of $+,-$ and a zero-sequence diagonal term.} & {Frequency interpretation requires specified channels; not a passivity test.} \\
	\bottomrule
	\end{tabular}
	\end{table}

	At the signal level, the contribution is structural rather than numerical: the usual sequence components become orthogonal idempotent projections in $\mathbb{R}^6$, with a direct Parseval identity and local $\Cl_{2,0}(\mathbb{R})$ phase planes. The multi-harmonic extension keeps the same projector language in $\mathbb{R}^{6h}$, which is useful when only selected harmonic--sequence components are required, for example in active-filter reference generation.

	This analytic-coordinate viewpoint should be distinguished from established time-domain extensions such as instantaneous symmetrical components and generalized symmetrical components, both of which address time-varying or nonsinusoidal waveforms directly~\cite{chicco2019}. The present work does not attempt to replace those formulations. Its role is to expose the real spectral, metric and rotor structure of the phasor/analytic representation; a full treatment of residual time-domain components, windowed estimators and nonstationary waveforms is left as a separate problem.

	The operator level relates two different symmetries. Modal blocks measure sequence coupling, while Cartan/eigenblade blocks resolve the fixed, half-turn and quadrature sectors of the Fourier alignment. For complex-linear operators, absence of off-diagonal Cartan blocks is equivalent to $\bm Q$-equivariance; a general real operator must additionally commute with $\mathcal J$ inside the quadrature block. The admittance example shows that the modal and Fourier diagnostics are independent: a balanced admittance can be diagonal in sequence blocks without being diagonal in eigenblade blocks, and a $\bm Q$-equivariant average can retain modal off-diagonal blocks. Equation~\eqref{eq:nested-residual} further separates physical phase interchange from the remaining restriction associated with the chosen Fourier frame.

Equivariance with respect to the quarter-turn $\mathcal J$ is exactly complex linearity. Proposition~\ref{prop:best-approx} characterises $\rho_{\mathcal J}$ as the minimum relative error of that approximation, with the metric and complex structure specified. Theorem~\ref{thm:splitting} identifies the allowed conjugate blocks under cyclic symmetry, including the zero-sequence exception. This is an operator statement: statistical impropriety additionally depends on the input covariance~\cite{picinbono1995,schreier2010}. A physical mirror-frequency interpretation requires explicit channels, as in the synchronous-frame derivation of Section~\ref{sec:ex-mirror} and the converter models of~\cite{harnefors2007complex,rygg2016modified,bakhshizadeh2016}. The residual does not establish passivity or stability; the dissipative $dq$ example has nonzero mirror coupling. Connection to converter stability therefore requires the device dynamics and network model~\cite{harnefors2007input}.

The machine example in Section~\ref{sec:ex-machine} supplies an observable consequence of this distinction: the orthogonal projection changes the voltage required for torque-ripple compensation. It extends the elementary frequency-pair model to a nonlinear magnetic benchmark while preserving the interpretation of $\rho_{J_4}$ as a local operator residual. It does not establish a new compensation law or a numerical advantage of rotor evaluation over an equivalent complete classical representation.

The sparse $\mathsf P/\mathsf N$ factorisation of Remark~\ref{rem:PN-plane-transfer}, following a phasor Clarke rotation~\cite{clarke}, provides an implementation counterpart to the rotor construction. Supplementary Section~S2 makes the coordinate transfers and the recovery of the invariant sequence projectors explicit. This also indicates a route to sparse multiphase implementations, provided local plane orientation is kept distinct from the global complex/conjugate splitting.

	The same caution applies to interharmonics. Formally, each analysed frequency bin could be assigned its own copy of $\mathbb{R}^6$ and its own modal projector triple; however, standard measurement windows, grouping rules and consistency conditions for harmonics and interharmonics introduce signal-processing constraints that are not addressed here~\cite{chicco2019,chicco2017unbalance}. In particular, once unbalance and distortion coexist, the assignment of measured spectral content to sequence components depends on the grouping convention adopted, so the indices computed downstream are not uniquely determined by the algebra alone~\cite{chicco2017unbalance}. The direct-sum construction should therefore be read as the algebraic backbone on top of which such estimators may be placed, not as a complete measurement protocol.

	\section{Conclusion}\label{sec:conclusion}

	We have reformulated Fortescue's symmetrical components as a real spectral decomposition in $\Cl_{6,0}(\mathbb{R})$. Realified phasors and Hilbert-lifted measurements lead to the same modal planes, jointly determined by the cyclic phase shift and the internal complex structure. Their projectors expose the metric, Parseval and local phase-plane structures. The worked examples verify that classical Fortescue coordinates, real modal projections and the inverse rotor sandwich agree to numerical precision.

	The closed frame-alignment rotor and its Cartan factorisation connect the modal construction to the three-point discrete Fourier transform. {The Fourier eigenstructure explains the seeded reconstruction, the phase-interchange symmetry and the distinct periods of the fractional path: eight for the spin lift and four for its vector action.} Modal and Fourier operator blocks diagnose independent symmetries.  {The complex/conjugate splitting adds a separate approximation criterion: $\rho_{\mathcal J}$ is the optimal relative complex-linear approximation error. Cyclic symmetry restricts inter-sequence coupling to conjugate exchange of positive and negative sequence, while allowing a zero-sequence diagonal contribution. The $dq$ example derives the mirror-frequency interpretation from explicit dynamics and demonstrates that coupling can coexist with passivity.}

	The same projector language extends directly to stacked harmonic spaces $\mathbb{R}^{6h}$, where any harmonic--sequence component is selected by a single idempotent and the classical harmonic rule modulo three is recovered as the balanced special case. The sparse $\mathsf{P}/\mathsf{N}$ plane-transfer form gives a complementary coordinate-level implementation of the same modal transform. Practical use requires the estimation and modelling choices discussed in Section~\ref{sec:discussion}. Future work should focus on realified impedance and admittance operators, converter models, faulted networks, and time-windowed or interharmonic extensions of the present framework{, and on the $N$-phase Fourier rotor in $\mathrm{Spin}(2N)$, whose Cartan structure is dictated by the Fourier eigenvalue multiplicities}.

	The nonlinear PMSM example gives the approximation criterion a concrete modelling consequence. For the selected 2\% voltage-unbalance case, the full local operator yields 96.55\% torque-ripple RMS reduction, compared with 42.39\% for its optimal complex-linear projection, using less corrective voltage. Independent nonlinear validation therefore complements the representation identities: equivalent coordinates preserve the response, whereas restricting the operator can change an engineering decision.

\section*{Declaration of competing interest}
	The authors declare that they have no competing financial or non-financial interests, currently or previously, related to this manuscript. Neither author currently serves, or has previously served, as an editor or editorial board member of Mathematics and Computers in Simulation.

	\section*{Data availability}
The algebraic examples use the models and parameters specified in the article. The nonlinear machine example uses the public MathWorks model cited in Section~\ref{sec:ex-machine}, whose original files are obtained through a licensed MATLAB/Simscape Electrical installation. Machine-identification and validation scripts, extracted operators, numerical records and figure-generation code accompany the article. The scripts reproducing Section~\ref{sec:examples}, the structural checks and the time-domain verification are supplied as supplementary material, together with dependency versions and execution instructions. No external experimental dataset was used.

\section*{Funding}
The authors received no specific funding for this work.

\section*{CRediT authorship contribution statement}
\textbf{Francisco G. Montoya:} Conceptualization, Methodology, Formal analysis, Software, Writing -- original draft, Writing -- review \& editing. \textbf{Ahmad H. Eid:} Methodology, Software, Validation, Writing -- review \& editing.

\section*{Declaration of generative AI and AI-assisted technologies in the manuscript preparation process}
During preparation of this submission, Codex (OpenAI) was used to assist with language editing, adaptation to the journal, bibliographic checks, simulation-scenario exploration, development and execution of simulation and analysis scripts, preparation of figures, and drafting of the numerical example and submission files. The authors are responsible for reviewing this assistance and for the accuracy and integrity of the manuscript.

\bibliographystyle{elsarticle-num}
\bibliography{references_MATCOM}

\end{document}